\documentclass[10pt,preprint2]{aastex}

\usepackage{natbib}
\usepackage{graphicx}

\usepackage{graphicx,caption,subcaption}
\usepackage{amsmath}
\usepackage[utf8]{inputenc}
\usepackage [english]{babel}
\usepackage [autostyle, english = american]{csquotes}
\MakeOuterQuote{"}

\usepackage{epstopdf}

\shorttitle{Photoionization Models of the Inner Gaseous Disk of the Herbig Be Star BD+65\,1637}
\shortauthors{Patel et al.}

\begin{document}

\title{Photoionization Models of the Inner Gaseous Disk of the Herbig Be Star BD+65\,1637}

\author{P. Patel\altaffilmark{1}, T. A. A. Sigut\altaffilmark{1} and J. D. Landstreet\altaffilmark{2} \\ {\small \it Accepted for publication in The Astrophysical Journal}}
\affil{Department of Physics and Astronomy, The University of Western Ontario,\\ London, Ontario Canada N6A 3K7}
\email{ppatel54@uwo.ca}
\altaffiltext{1}{Center For Planetary Science \& Exploration, The University of Western Ontario, London, Ontario N6A 3K7}
\altaffiltext{2}{Armagh Observatory, Armagh, Northern Ireland, UK}

\begin{abstract}
We attempt to constrain the physical properties of the inner, gaseous disk of the Herbig Be star BD+65\,1637 using non-LTE, circumstellar disk codes and observed spectra (3700 to 10,500~\AA) from the ESPaDOnS instrument on CFHT. The photoionizing radiation of the central star is assumed to be the \textit{sole} source of input energy for the disk. We model optical and near-infrared emission lines that are thought to form in this region using standard techniques that have been successful in modeling the spectra of Classical Be stars. By comparing synthetic line profiles of hydrogen, helium, iron and calcium with the observed line profiles, we try to constrain the geometry, density structure, and kinematics of the gaseous disk. Reasonable matches have been found for all line profiles individually; however, no disk density model based on a single power-law for the equatorial density was able to simultaneously fit all of the observed emission lines. Amongst the emission lines, the metal lines, especially the Ca\,{\sc ii} IR triplet, seem to require higher disk densities than the other lines. Excluding the Ca\,{\sc ii} lines, a model in which the equatorial disk density falls as $10^{-10} (R_{*}/R)^{3}~\rm g\,cm^{-3}$ seen at an inclination of $45^{\circ}$ for a $50\,$R$_{*}$ disk provides reasonable matches to the overall line shapes and strengths. The Ca\,{\sc ii} lines seem to require a shallower drop off as $10^{-10} (R_{*}/R)^{2}~\rm g\,cm^{-3}$ to match their strength. More complex disk density models are likely required to refine the match to the BD+65\,1637 spectrum. 
\end{abstract}

\keywords{stars: pre-main sequence, stars: variable: T Tauri, Herbig Ae/Be, stars: individual (BD+65\,1637), stars: emission, Be, accretion disks, line: profiles}

\section{Introduction}
Herbig Ae/Be (hereafter HAeBe) stars are pre-main sequence A or B-type stars with emission lines and an excess in their infrared spectral energy distributions (SEDs). The emission lines, particularly the Balmer series of hydrogen, and the infrared excess can be attributed to circumstellar dust and gas, the likely remnants of the star formation phase~\citep{Herbig1960, FM1984, WW1998}. Circumstellar dust distinguishes HAeBe stars from the Classical Be stars,\footnote{Classical Be stars are B-type, main-sequence stars that show, or have once shown, one or more Balmer emission lines in their spectrum.} whose infrared excess is solely due to free-free emission from the ionized, dust-free gas in a circumstellar decretion disk~\citep{Rivinius2013}.

Being the precursors to the debris disks, such as those around $\beta$ Pictoris and Vega, HAeBe stars make interesting subjects for studying disk physics, as well as for understanding disk evolution in pre-main sequence stars~\citep{PG1997}. HAeBe stars are also an important link between low and high mass star formation. High mass, O-type stars form at the centers of very dense clusters and involve complex environments. The formation of such stars is currently poorly understood~\citep{Larson2003}. In addition, such stars spend their pre-main sequence life in a deeply embedded state before becoming optically visible as main-sequence objects, hence depriving us of the opportunity to observe the early phase of star formation~\citep{ZY2007}. In contrast, Herbig Be stars (HBe hereafter), despite forming in complex and dense environments, may become optically visible just before they reach the Zero Age Main Sequence (ZAMS) as they spend comparatively longer in the pre-main sequence (PMS, hereafter) phase compared to their higher mass counterparts. This not only aids in the understanding of the formation process of intermediate mass HBe stars, but can also help to bridge the understanding of the star formation process between low mass T Tauri stars\footnote{T Tauri stars are pre-main sequence stars, with masses less than 2.5\,M$_{\sun}$, which show Balmer emission lines in their spectrum and an excess in their infrared SEDs, interpreted as gas and dust in form of a circumstellar disk. The most massive T Tauri stars will later become main-sequence A-type stars.} and high mass, O-type stars.

Knowledge of the physical conditions in the disks around HAeBe stars has been steadily increasing. The disk material is inherited by the star from its parent molecular cloud. Near-IR and millimeter interferometric observations (see the review  by~\cite{Kraus2015} and references therein) have been instrumental in providing strong evidence for circumstellar disks around HAeBe stars. The disk can extend to 100s of AUs and disk temperatures can vary from a few 10s K to 1000s K~\citep{DM2010}. These disks have been studied extensively at far-infrared and millimeter wavelengths for cooler dust species such as PAHs (Polycyclic Aromatic Hydrocarbons), iron oxide grains, and silicates (crystalline and amorphous) to understand the evolution of these species in the disk~\citep{WW1998}. Near and mid-infrared wavelengths have been used to study molecular gas and warm dust in regions closer to the star. 

\cite{Vink2002},~\cite{Mottram2007}, and~\cite{Vink2015} have shown that there exists a difference in polarization between Herbig Ae (hereafter HAe) and HBe stars, especially early B-type stars.~\cite{Vink2002} first noticed that while all HAe stars show intrinsic linear polarization consistent with magnetospheric accretion, most HBe stars show line-depolarization which is consistent with disk accretion. HAeBe stars show photometric variability ranging from days to months to years~\citep{Herbig1960,FM1984}, and various models such as non-radial pulsation, accretion, etc, have been suggested for their occurrence~\citep{Catala1994}. ~\cite{vandenAncker1998} and~\cite{Mendigutia2011} have shown that HAe stars show large to moderate variations in magnitude ($>$2$^{m}$.5) while HBe stars show low to moderate variations ($<$0$^{m}$.5). The differences between the variations has been suggested to be due to different accretion mechanisms. In addition,~\cite{CJK2014} studied He\,{\sc i} ($\lambda\,10830$)\footnote{All wavelengths are in~\AA~ in this study unless otherwise stated.} in a sample of HAeBe stars and noticed that HBe stars show blue shifted absorption features while HAe stars show both blue and red shifted absorption features. This difference indicates that HBe stars show little evidence of infalling material, while HAe's show a higher level of mass flow activity, suggesting the action of different mechanisms. A recent study by~\cite{Fairlamb2015} of UV excess of a large set of HAeBe stars showed that early-type HBe stars cannot be modeled successfully using the magnetospheric accretion, again hinting of an alternate mode of accretion. 

~\cite{Alecian2008,Alecian2013} studied a large set of HAeBe stars and found that less than 10\% of them show evidence of large scale magnetic fields. Possible causes for a lack of large scale magnetic fields for the HAe and HBe stars are either a very small convective core or completely radiative envelope, neither of which generate large scale magnetic fields~\citep{Alecian2014}. Accretion via a magnetosphere is fairly well established for T Tauri stars (see reviews by ~\cite{Bouvier2007}~and~\cite{GomezDeCastro2013}) but whether or not the same mechanism applies to HAeBe stars is still an open question. With the differences in the polarizations and the lack of large scale magnetic fields, the mechanism at work for accretion in HBe stars is still a mystery.   

\section{Disk Structure}
\subsection{Atomic Gaseous Disk}
\label{sec:gas_disk}

Close to the star, one expects regions with high temperatures that would destroy any dust and create a dust-free zone. Further away, the temperature is cool enough for dust to exist, and in the region favorable for dust, one either finds a thick wall of dust, sometimes called a dust rim, or a smooth transition to dusty disk at the dust sublimation radius (see~\cite{DM2010} for a review).~\cite{Monnier2005} suggest that if an optically thin gas exists in the dust-free zone, a wall of dust is expected as direct radiation from the star will enhance the scale height of the dust rim or puff up the inner disk wall. If an optically thick gas exists, then the transition to dust is smoother. 

\begin{figure*}
\centering
\epsscale{2.0}
\plotone{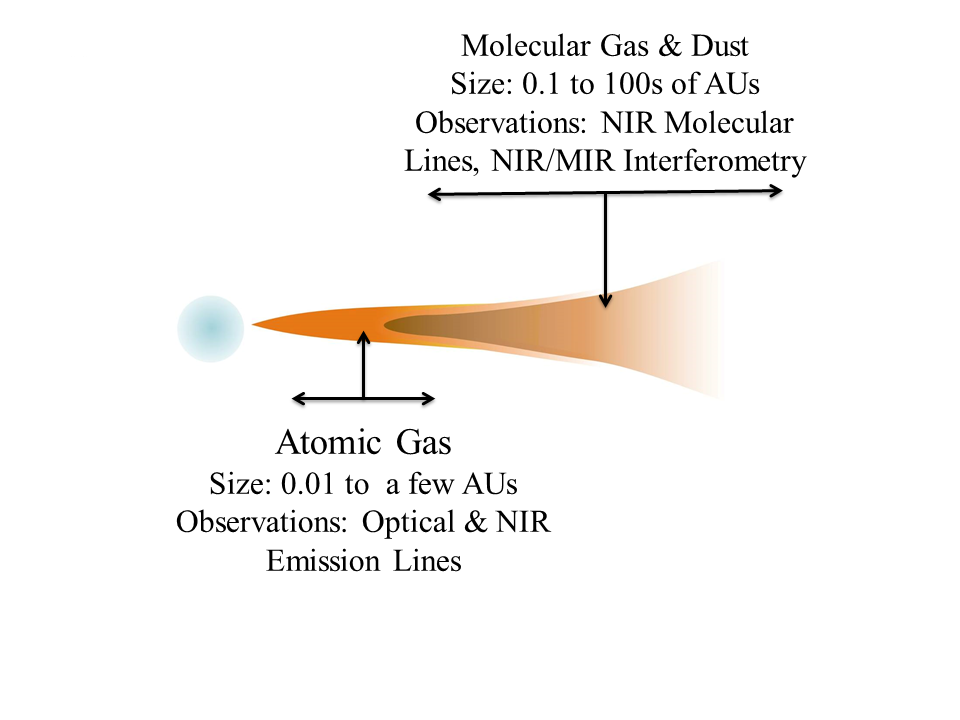}
\caption{Inferred structure of the disk around HBe stars. Close to the star, partially ionized, atomic gas is expected. The size of this region may extend anywhere from 0.01 AU to a few AUs. This region can be studied using optical and near-infrared spectral lines. As the temperatures cool in the disk, molecular gas and dust is expected. The size of this outer disk ranges from 0.1 to 100s of AUs. The transition between atomic and molecular gas occurs at around a few AUs. The molecular gas and dust found beyond the atomic gaseous disk is represented by brown, and this region can be studied using variety of different methods such as molecular lines and NIR/MIR interferometry. It is important to note that this image is not to scale. Image Credit: Ami Patel}
\label{fig:YSOdiskstructure}
\end{figure*}

As dust evaporates above $\sim$1500K, the portion of the disk closest to the star is likely to be completely gaseous due to high temperatures resulting from the star's UV radiation. Depending on the distance from the central star and the stellar radiation, HBe disks can be divided into atomic gas, molecular gas, and dust. The expected structure of such a disk is illustrated in Figure~\ref{fig:YSOdiskstructure}. The inner gaseous region extends from, perhaps, 0.1 AU to a few AUs ($\sim100R_{*}$) in size, and temperatures can reach to 1000s K. Optical and Near-IR (NIR, hereafter) atomic lines are used to study this hot, atomic gaseous region. The disk region beyond the gaseous disk receiving direct radiation from the star are expected to have a thin surface layer with atomic gas; hence, the emission seen in the spectrum can arise from a large extended area. In the region beyond the atomic gaseous disk, where the temperatures are cooler, molecular gas is expected. As the temperature decreases, the molecular gas condenses into dust and a mixture of warm dust and molecular gas is expected. NIR molecular emission lines and NIR/Mid-IR (MIR, hereafter) interferometry is used to understand the structure of this region. Beyond this region, cooler gas and dust are detected using millimeter and submillimeter interferometry. The region beyond the atomic gaseous disk can range anywhere in size from 0.1 to 100s of AUs, including the dusty disk. The outer, dusty region of the disk is well studied, but currently very little is known about the inner, gaseous region.

Several studies have suggested that HBe stars can be compared to Classical Be stars (see~\cite{HP1992,Mottram2007}) in their circumstellar disk and  structure.~\cite{Hillenbrand1992} noticed that some of the HAeBe stars in their sample show a small IR excess, comparable to that seen in Classical Be stars. Like early-type HBe stars, Classical Be stars are known to show depolarization (see~\cite{Rivinius2013}). In addition, similarities such as the H$\alpha$ equivalent width distribution, fast rotation speeds (i.e. high $v\,sini$), photometric variability, and slow outflow velocities further connect the Classical Be and HBe stars~\citep{BB2000}. Hence, using models that have been well established for Classical Be stars would make a good first step in understanding the inner, atomic gaseous disks of HBe stars.

\subsection{Molecular Gas \& Dust Disk}

Several studies of molecular gas using tracers such as Br$\gamma$ (~\cite{Mendigutia2011,Calvet2004}) and CO (e.g.~\cite{Ilee2014,Ilee2013,Wheelwright2010}) have been conducted, probing the relatively warm region of the disk where these emission lines are thought to originate.~\cite{Ilee2014} studied CO overtone emission and strongly suggests that the emission originates from a small gaseous disk inside the dust sublimation radius that follows Keplerian rotation. This hints at the process of classical disk accretion where material is thought to be transfered from the disk directly onto the star through an equatorial disk.

~\cite{MMG2002} and~\cite{Millan-Gabet2007} noticed that most HBe stars have a smaller dust sublimation radius ($R_{rim}$) than predicted by the luminosity-size ($R_{rim} \propto L_{*} ^{1/2}$) relationship developed for the T Tauri and HAe stars. ~\cite{MMG2002} suggest that the smaller size of the gaseous disk around HBe stars may be due to optically thick gas absorbing the UV radiation, allowing the dust to exist closer to the star.~\cite{Eisner2004} was able to fit an inner rim model to a flat disk with optically thick gas to 2.2 $\mu m$ observations for higher mass stars.~\cite{Alonso-Albi2009} showed that the disks around HBe stars are 5-10x less massive than those of lower mass counterpart HAe and T Tauri stars and proposed that strong UV radiation from the hot, central star evaporates and disperses the gas, leaving behind a small dusty disk. Finally, the gas in the circumstellar disk is seen to follow Keplerian rotation, as found by~\cite{MS1997,MS2000} using millimeter interferometric measurements.

\subsection{Emission Lines}
\label{sec:emissionlines}

By definition, all HAeBe stars show emission lines in their spectra which can be used to trace the structure and processes in the disk.~\cite{HP1992} show that the spectra of HAeBe stars are very similar to those of T Tauri stars; however, they note that the similarity of the spectra does not mean that the formation mechanism of the lines is the same. The higher stellar temperatures of the HAeBe stars can change both the mechanisms and the extent to which the gas is excited due to the intense stellar radiation field experienced by the disk gas.  

Only a few studies have qualitatively modeled the permitted emission lines seen the visible and NIR part of the spectrum.~\cite{CK1979},~\cite{FM1984},~\cite{HP1992} and~\cite{BC1995} investigated emission lines such as H$\alpha$, the Ca\,{\sc ii} IR triplet ($\lambda\,8498$, $\lambda\,8542$ \& $\lambda\,8662$) and several Fe\,{\sc ii} lines and interpreted them as revealing chromospheric and wind activity in these stars. More recently, He\,{\sc i} ($\lambda\,10830$) has been used as a tracer of mass flow activity by~\cite{CJK2014}. In addition to permitted emission lines, forbidden emission lines are found to be present in HAeBe stars, and~\cite{CR1997} have shown that these lines may arise from the winds and outflows.

~\cite{FM1984} showed that the H$\alpha$ line in HAeBe stars can be divided into three groups based on the line profile morphology, namely single-peaked, double-peaked and P-Cygni line profiles.  More than 50\% of the stars studied showed a double-peaked H$\alpha$ profile, while the rest were divided equally into single-peaked and P-Cygni profiles.~\cite{FJ1984} note that all Balmer lines exhibit the same line shape as H$\alpha$, with the strength of the line decreasing from H$\alpha$ to H$\gamma$. H$\alpha$ has been widely used for wind  diagnostics studies of Herbig Ae/Be stars~\citep{FM1984,CR1998} and polarization~\citep{Vink2002,Vink2015}. 

~\cite{HP1992} and~\cite{HP1992b} investigated the Ca\,{\sc ii} triplet emission in HAeBe stars: 71\% of their HBe stars showed the Ca\,{\sc ii} IR triplet lines in emission. They showed that if the excitation is same in all the stars, Ca\,{\sc ii} requires denser and/or thicker regions in hotter stars (see their Figure 8). They also noted that due to high stellar temperatures and the double-peaked line profiles, the formation of the Ca\,{\sc ii} lines would happen away from stellar surface in a disk like those possessed by Classical Be stars and in a very small, ring-like structure close to the star. They also showed that there is a correlation between Ca\,{\sc ii} luminosity and IR excess in these stars and concluded that Ca\,{\sc ii} emission lines are somehow related to the presence of \textbf{a} disk. 

~\cite{BC1995} studied non-photospheric lines such as the H$\alpha$, the Ca\,{\sc ii} IR triplet, and He\,{\sc i} ($\lambda\,5876$). They concluded that the energy fluxes for these lines increase with the effective temperature and suggest that the origin of the emission in the lines is further away from the layer between the disk and the stellar surface.~\cite{Hernandez2004} found that 33\% of their sample showed emission in Fe\,{\sc ii} multiplets and also noticed that the equivalent width (EW, hereafter) of one of the Fe\,{\sc ii} lines, Fe\,{\sc ii} multiplet (42) ($\lambda\,5169$), is correlated with the equivalent width of H$\alpha$ and seems to be present only when [O\,{\sc i}] ($\lambda\,6300$) also appear in the spectrum. 

\subsection{Overview}

Differences in polarization, magnetism, and spectral lines between HAe and HBe stars suggest the possibility of a different accretion mechanisms at work. The accretion mechanism may be constrained by studying material very close to the star, determining its structure and kinematics. As high temperatures are expected close to the star, we chose emission lines that form outside the stellar photosphere and require high temperatures to be excited. Modeling these lines and investigating what kind of disk density structure they require will allow us to qualitatively understand the regions close to the star. 

In this work, we model emission lines thought to be produced in the inner gaseous disk of one HBe star, BD+65\,1637, using observations from CFHT ESPaDOnS. Section~\ref{sec:star_info} describes the observations, reduction methods, and the emission lines found in the spectrum. The details of  the models can be found in Section~\ref{sec:modeling}, and results of the modeling for each emission line can be found in Section~\ref{sec:results}. The uniqueness of the models is discussed in Section~\ref{sec:uniquediskmodels}. The paper concludes with a discussion in Section~\ref{sec:discussion}, and a summary of key findings in Section~\ref{sec:conclusions}.

\section{BD+65\,1637}
\label{sec:star_info}

BD+65\,1637 (V*\,V361 Cep) is a B2e star of visual magnitude 10.83 in the young cluster NGC 7129 ~\citep{straizys2013,DH2015}. BD+65\,1637 was identified as HBe star by George Herbig in his first paper on Herbig stars~\citep{Herbig1960} and noted to a have spectrum very much like that of a Classical Be star.~\cite{Hillenbrand1992} studied the star's spectral energy distribution (SED) which showed a small infrared excess, and they classified it  as very similar to Classical Be stars. The assigned spectral type has varied in the literature from B2 to B5 (see \cite{Herbig1960,strom1972,FM1984,fink1985,hillenbrand1995}). Here, we will adopt the stellar parameters and their uncertainties from~\cite{Alecian2013}, reproduced in Table~\ref{table1}.

\begin{table}[h]
\caption{Stellar parameters for BD+65\,1637}
\label{table1}
\smallskip
\begin{center}{\small
\resizebox{0.45\textwidth}{!}{
\begin{tabular}{ll} 
\hline \hline
\noalign{\smallskip}
Parameter & Value\\
\hline
 \noalign{\smallskip}
Spectral Type & B2e\\
\noalign{\smallskip}
T$_{\rm eff}$ (K) & 18000$\pm$1000\\
\noalign{\smallskip}
log \textit{g} (cgs) & 4.0\\
\noalign{\smallskip}
Radius  (R$_{\sun}$) & 6.7$\pm$0.7\\
\noalign{\smallskip}
Mass (M$_{\sun}$) & 8.11$^{+0.24}_{-0.23}$\\
\noalign{\smallskip}
Distance (pc) & 1250$\pm$50\\
%\noalign{\smallskip}
%Age (Myr) & 0.035$^{+0.012}_{-0.010}$\\
\noalign{\smallskip}
$v\,sini$ ($\rm km\,s^{-1}$) & 278$\pm$27\\
\noalign{\smallskip}
\hline 
%\hline
\noalign{\smallskip}
\noalign{\smallskip}
	\begin{minipage}{6 cm}
	All the values for the parameters as well as their uncertainties are taken from~\cite{Alecian2013}.
	%\textbf{Note: The uncertainty for $T_{eff}$ mentioned here is result of fitting models to the wings of the Balmer lines and visual inspection of the fits by fixing a value to \textit{log g}. Thus, the uncertainty for \textit{log g} is $\pm$0.5 dex.}
	\end{minipage}
\end{tabular}}}
\end{center}
\end{table}

\subsection{Observations}
\label{sec:observations}

The observational data were obtained in 2006 (HJD-2453898) using the high-resolution spectropolarimeter ESPaDOnS at the Canada-France-Hawaii Telescope. Additional spectra are also available from the Narval spectropolarimeter at the T\'{e}lescope Bernard Lyot, obtained in 2009 (HJD-2455099). Both spectropolarimeters cover the wavelength range from 3700 to 10500~\AA~ with a spectral resolution of 65,000. The peak SNR per CCD pixel at 7300~\AA\ was 237 for ESPaDOnS spectrum and 276 for Narval spectrum.

\begin{figure*}
\epsscale{2.3}
\centering
\plotone{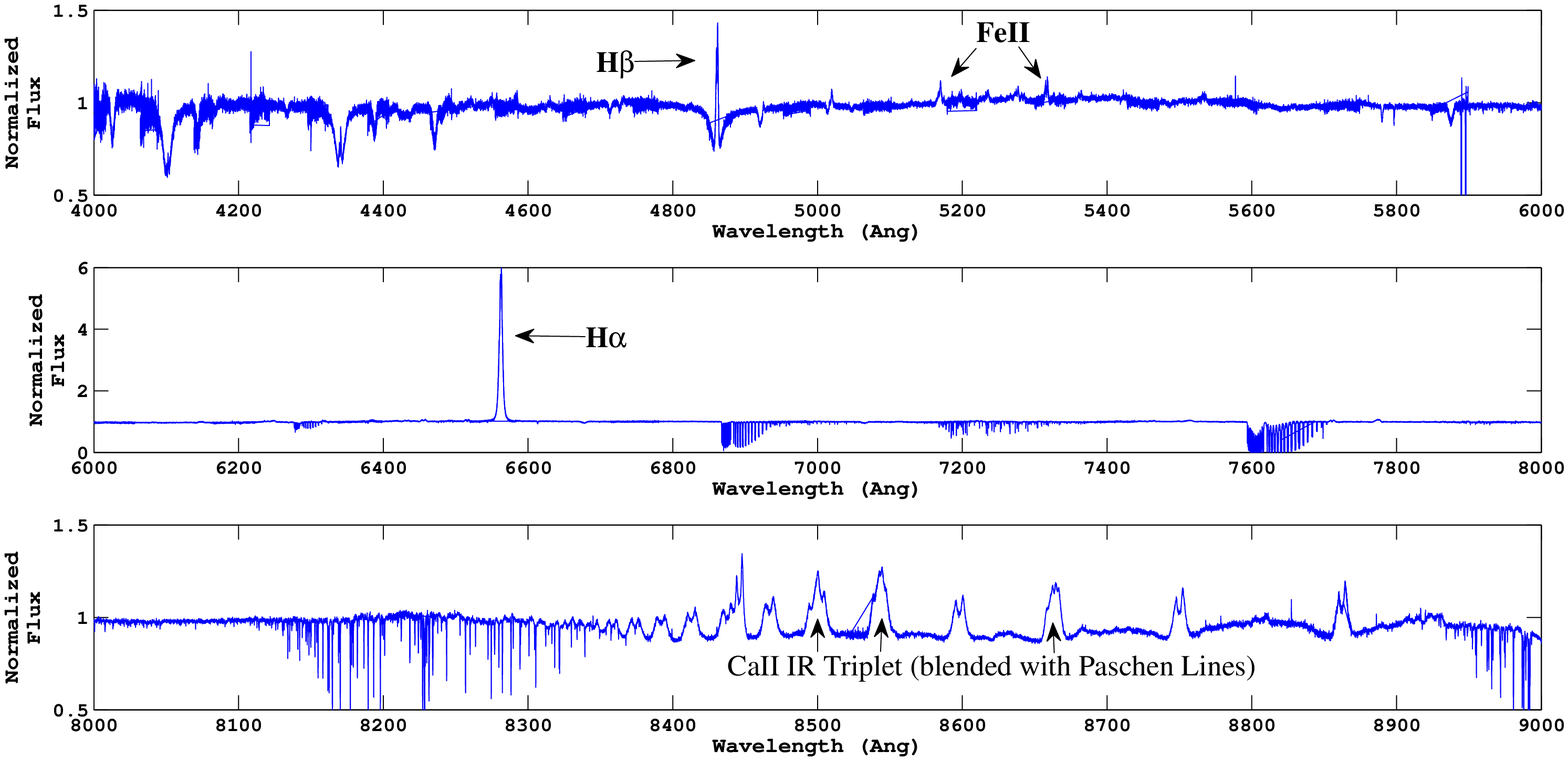}
\caption{The observed spectrum for BD+65\,1637~\citep{Alecian2013}. The spectrum shows many emission lines, such as the hydrogen Balmer lines, H$\alpha$ ($\lambda\,6563$) and H$\beta$ ($\lambda\,4861$), the Ca\,{\sc ii} IR-triplet lines ($\lambda\,8498$, $\lambda\,8542$ and $\lambda\,8662$) and the Fe\,{\sc ii} lines ($\lambda\,5169$ of multiplet 42 and $\lambda\,5317$ of multiplet 49). The strength of the lines vary over a large scale, particularly for H$\alpha$ as seen in the vertical scale of each subplot.}
\label{fig:bd_spec}
\end{figure*}

\begin{figure*}
\centering
\epsscale{2.3}
\plotone{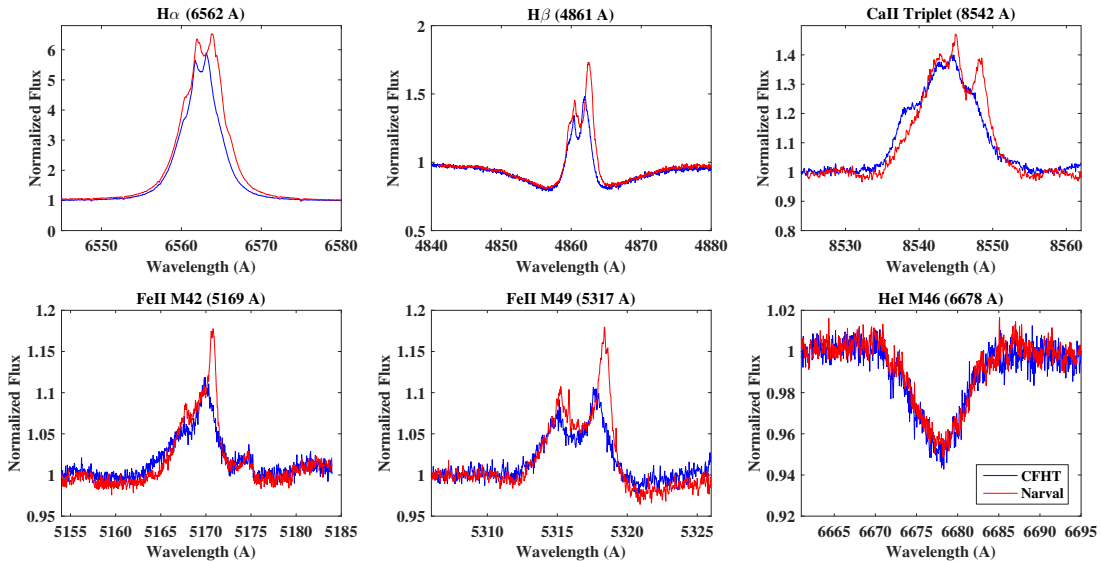}
\caption{The CFHT ESPaDOnS observations (blue) are over-plotted with Narval observations (red) for each individual line. The ESPaDOnS observation was obtained in 2006 (HJD-2453898) at Canada-France-Hawaii Telescope, and the Narval observations were taken in 2009 (HJD-2455099) at  the T\'{e}lescope Bernard Lyot. Except for He\,{\sc i} ($\lambda\,6678$), all the lines show variations with time.}
\label{fig:cfht_narval}
\end{figure*}

BD+65\,1637's CFHT spectrum can be seen in Figure~\ref{fig:bd_spec}. The spectrum not only contains strong Balmer line emission, H$\alpha$ and H$\beta$, but also emission in many metal lines such as those from calcium, oxygen and iron. In addition to the two strong Balmer lines, we will investigate one of the three Ca\,{\sc ii} infrared triplet lines ($\lambda\,8542$), two Fe\,{\sc ii} lines ($\lambda\,5169$ of multiplet 42 and $\lambda\,5317$ of multiplet 49) and one He\,{\sc i} line ($\lambda\,6678$). The He\,{\sc i} line is seen in absorption and is used to estimate stellar properties such as the \textit{$v\,sini$} of the star. A detailed profile of each line can be found in Figure~\ref{fig:cfht_narval}. 

%\begin{figure*}
%\centering
%\epsscale{2.3}
%\plotone{Figure3.eps}
%\caption{The observed individual emission lines for BD+65\,1637. The hydrogen Balmer lines, H$\alpha$ ($\lambda\,6562$) and H$\beta$ ($\lambda\,4861$), one of the Ca\,{\sc ii} IR-triplet lines ($\lambda\,8542$), Fe\,{\sc ii} multiplets ($\lambda\,5169$ and $\lambda\,5317$). A photospheric line, He\,{\sc i} ($\lambda\,6678$) is also included in this study. Note: The scale of the y-axis varies from panel to panel.}
%\label{fig:bd_emissionlines}
%\end{figure*}

The Balmer lines, H$\alpha$ and H$\beta$, are the strongest emission lines in the optical/NIR spectrum of BD+65\,1637, as illustrated in Figure~\ref{fig:bd_spec}. For BD+65\,1637, the equivalent width of the H$\alpha$ lines has been noted to vary from -45~\AA~to -28~\AA\ (\cite{Hernandez2004},~\cite{Fernandez1995},~\cite{FM1984} and~\cite{GA1977}). In the 2006 CFHT spectrum, the EW measured for H$\alpha$ is -26.2~\AA\ and for H$\beta$, -0.92~\AA. The EW for Ca\,{\sc ii} IR Triplet ($\lambda\,8542$) is measured to be -4.0~\AA. The two Fe\,{\sc ii} lines, $\lambda\,5169$ and $\lambda\,5317$, were chosen for this study as they are in different multiplets, and their EWs were measured to be -0.58~\AA~and -0.40~\AA. Fe\,{\sc ii} $\lambda\,5169$ is in multiplet 42 ($a ^{6}S - z ^{6}P^{\circ}$) and Fe\,{\sc ii} $\lambda\,5317$ is in multiplet 49 ($a ^{4}G - z^{4}F^{\circ}$), although it is blended with a weaker line from multiplet 48.\footnote{While both multiplets arise from the $a^{4}G$ lower level, the $\Delta_{ji}$ value of the multiplet 48 component is 6.5 times smaller than the multiplet 49 component.} Both of these lines have low excitation energy. He\,{\sc i} ($\lambda\,6678$) is in absorption, and the EW was measured to be 0.33~\AA.

A comparison between the CFHT and Narval spectra, taken approximately three years apart, can be seen in Figure~\ref{fig:cfht_narval}. Some variation in the spectral lines is seen, most notably in the metal lines and V/R ratio, with a strong red component in Narval (2009) observations. The similarity in the He\,{\sc i} ($\lambda\,6678$) absorption line, in contrast to variations in the Balmer and metal lines, confirms that the Balmer and metal emission lines arise from material outside the star and that the disk structure varies over time. For the analysis of this paper, only the 2006 CFHT ESPaDOnS data was used. 

H$\alpha$, H$\beta$, and both the Fe\,{\sc ii} lines ($\lambda\,5169$ and $\lambda\,5317$) show double-peaked profiles, with a stronger red component (R) when compared to the blue component (V). All the non-photospheric lines included in this study show changes in the V/R ratio over time (see Figure~\ref{fig:cfht_narval}). Classical Be stars are known to show V/R line variability which is interpreted as caused by a one-armed, global disk oscillation (for details see the review by~\cite{Rivinius2013}). This effect has not been included in the model here, and hence the synthetic line profiles cannot fit both peaks of a line profile simultaneously. 

\subsection{Reduction of Spectra}
\label{sec:reduction_method}

In order to compare observed and modeled line profiles, the observed spectra needed to be continuum normalized. The unnormalized data obtained from CFHT (and Narval) were separated into specific wavelength windows that included the emission lines of interest. Each wavelength window was continuum normalized using IRAF.~\footnote{IRAF is distributed by the National Optical Astronomy Observatories, which are operated by the Association of Universities for Research in Astronomy, Inc., under cooperative agreement with the National Science Foundation.} The function and order used for normalization varied from one spectral window to another; `Legendre' and `cubic spline' functions and low-order polynomials were generally used in the process.  

\begin{figure*}
\centering
\epsscale{2.3}
\plotone{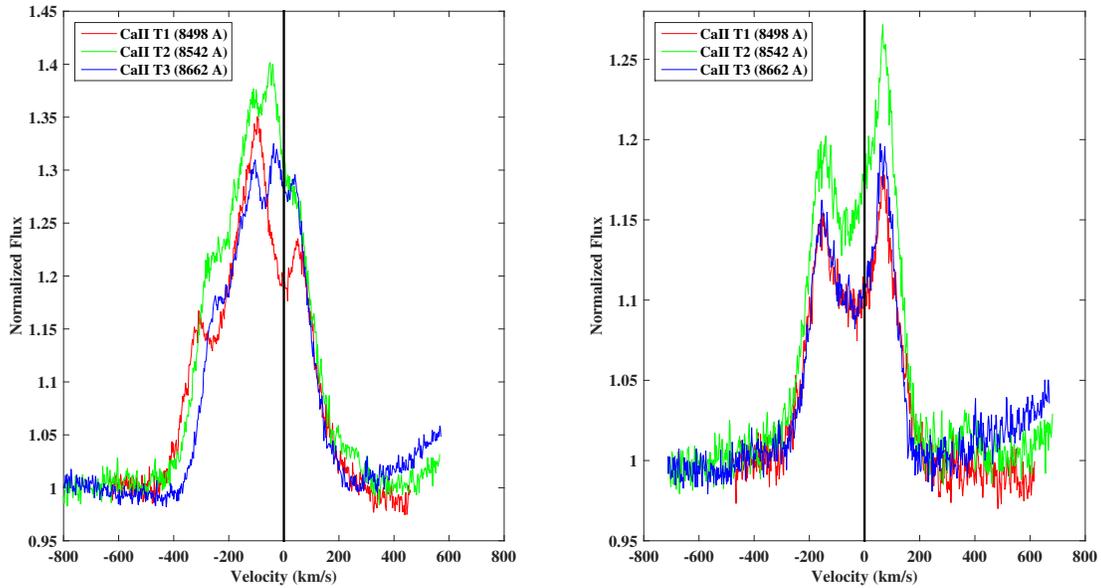}
\caption{\textit{Left:} All Ca\,{\sc ii} IR Triplets before the subtraction process (i.e. they are blended with Paschen lines). \textit{Right:} The Ca\,{\sc ii} IR Triplets after subtraction of Paschen lines. The black solid line denotes 0 km/s radial velocity.}
\label{fig:bd_caIIbeforeafter}
\end{figure*}

The Ca\,{\sc ii} infrared triplet lines ($\lambda\,8498$, $\lambda\,8542$ and $\lambda\,8662$) are blended with the hydrogen Paschen series. In order to compare with synthetic line profiles, the Paschen line must be subtracted from the Ca\,{\sc ii} line. This was done by taking an average of the two nearest, unblended Paschen lines and then subtracting the average from the Ca\,{\sc ii} line. Figure~\ref{fig:bd_caIIbeforeafter} illustrates all the three lines before the procedure of subtraction (on the left) and after the process of subtraction (on the right) in the CFHT spectrum. It can be seen that the subtraction process decreases the strength of the Ca\,{\sc ii} line, and the resultant line profile is double-peaked with a stronger red peak.

\begin{figure*}
\centering
\epsscale{2.3}
\plotone{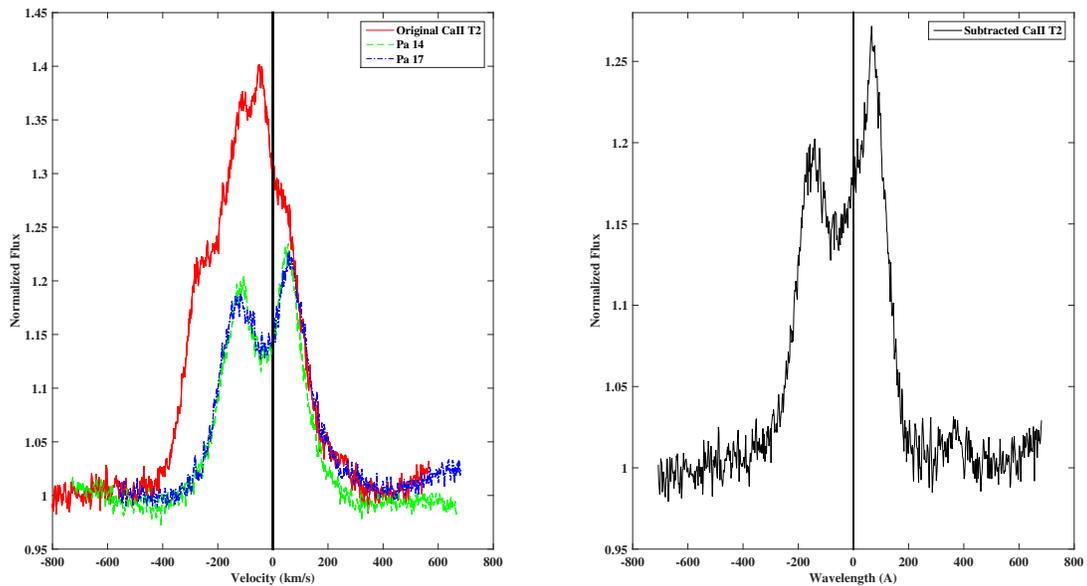}
\caption{\textit{Left}: One of the Ca\,{\sc ii} triplet lines ($\lambda\,8542$), seen in red, along with the Paschen 14 (P14, $\lambda\,8596$) in green and Paschen 17 (P17, $\lambda\,8467$) in blue. \textit{Right}: The resultant Ca\,{\sc ii} line profile after subtraction of the average of the Paschen lines. The different scales on the y-axis should be noted.}
\label{fig:bd_caIIcompare}
\end{figure*}

An example of the subtraction process is illustrated in Figure~\ref{fig:bd_caIIcompare}, where the average of Paschen 14 (P14, $\lambda\,8596$) and Paschen 17 (P17, $\lambda\,8467$) is subtracted from the blended Ca\,{\sc ii} $\lambda\,8542$ line to extract the unblended Ca\,{\sc ii} profile. For the rest of the paper, the resultant Ca\,{\sc ii} line profile (seen in the right hand side panel of Figure~\ref{fig:bd_caIIcompare}) will be used. All the lines are adjusted for the stellar radial velocity. The radial velocity required for the line shifts was measured using the center of He\,{\sc i} ($\lambda\,6678$), which was measured to be $-17.1~\rm km\,s^{-1}$ for the CFHT spectrum and $-4.1~\rm km\,s^{-1}$ for the Narval spectrum. Both the values are within the range of $-26\pm20~\rm km\,s^{-1}$ measured by~\cite{Alecian2013}.

\section{Modeling}
\label{sec:modeling}

To calculate the thermal structure of the equatorial, non-accreting, gaseous disk surrounding the central B~star, the {\sc Bedisk} code~\citep{SJ2007} was used. This code calculates the temperature structure of the disk, given a user-defined density structure, by enforcing radiative equilibrium in a gas of solar composition. \textit{The energy input into the disk was assumed to be solely from the photoionizing radiation field of the central star}. The microscopic rates of heating and cooling were balanced to determine the temperature at many grid points in the disk. The disk was assumed to be axisymmetric and in Keplerian rotation about the central star.\footnote{No wind from the central star has been included in the calculations.} {\sc Bedisk} includes nine abundant elements (H, He, C, N, O, Mg, Si, Ca and Fe) over many ionization stages in the determination of the radiative equilibrium temperatures. The atomic level populations, required for the calculation of the heating and cooling rates, as well as for use later in computing emission lines, were obtained by solving the statistical equilibrium equations in an escape probability formalism (see~\cite{SJ2007} for details).

The user defined density structure of the disk $(\rm g\,cm^{-3})$ was taken to be specified by the parameters $\rho_{0}$ (base disk density) and $n$ (radial power law index) in the equation
\begin{equation}
\rho(R,Z) = \rho_{0} \left(\frac{R_*}{R}\right)^{n} \,
e^{-\left(\frac{Z}{H}\right)^2} \;,
\label{eq:rho}
\end{equation}
where $R$ and $Z$ are the cylindrical co-ordinates for the disk, $R_{*}$ is the stellar radius, and $H$ is the disk scale height. In all our models, the disk extends from the stellar photosphere (R starts at R$_{*}$) out to a radius of R$_{disk}$.

If the vertical density structure of the disk is determined solely by gravitational equilibrium (the disk is rotationally supported in the $R$ direction), then the scale height has the form

\begin{equation}
H=\beta(T_{\rm HE}) \left(\frac{R}{R_{*}}\right)^{3/2},
\label{eq:scaleheight}
\end{equation}\\  
where
\begin{equation}
\beta(T_{\rm HE})=\sqrt{\frac{2kT_{\rm HE}R_{*}^{3}}{GM_{*}\mu M_{H}}} .
\label{eq:alpha}
\end{equation}\\
Here, $M_{*}$ is the mass of the star, $\mu$ is the mean molecular weight of the gas in the disk (taken to be 0.68), and $T_{\rm HE}$ is the hydrostatic equilibrium temperature assumed for the disk. This hydrostatic temperature is used solely for setting the vertical scale height of the disk, and typically in Classical Be stars, one takes $T_{\rm HE}\approx 0.6\,T_{\rm eff}$. A self-consistent treatment, in which the radiative equilibrium disk temperatures are used in the calculation of the vertical hydrostatic equilibrium, is possible (see~\cite{Sigut2009}), but that has not been used here. In the present work, varying $T_{\rm HE}$ allowed the scale height of the disk to be varied. Finally we note that vertical gravitational equilibrium predicts a flaring disk, $H\propto R^{3/2}$.

The assumption of vertical hydrostatic equilibrium produces a very thin disk near the star. From Eqn.~\ref{eq:alpha}, it is easy to see that the the ratio $H/R$ can be expressed as
\begin{equation}
\frac{H}{R}=\frac{c_{S}}{V_{K}},
\label{eq:scaleheightoverR}
\end{equation}\\  
where $c_{\rm S}$ is the local sound speed and $V_{\rm K}$ is the Keplerian orbital velocity. As the orbital velocity is typically hundreds of $\rm km\,s^{-1}$ while the sound speed is on the order of $10\,\rm km\,s^{-1}$ for the disk temperature, the disk is predicted to be geometrically thin with $H/R\ll 1$. Such thin disks have been very successful in interpreting observables in Classical Be stars,\footnote{There has been some evidence in the support of larger disk scale heights (see~\cite{SP2013,Arias2006,Zorec2007}).} but their application to HBe stars is less clear. For this reason, we also considered disk density models with an enhanced scale height, achieved by setting the $T_{\rm HE}$ temperature to be $5\,T_{\rm eff}$, a factor of nearly 10 over the gravitational equilibrium value. We will refer to these models as {\it thick disk\/} models. Note that for the same disk density parameters, $\rho_{0}$, $n$, and $R_{disk}$, the thick disk models are a factor of $\sqrt{5/0.6}$ more massive, as the total disk mass is proportional to the scale height.

Finally, we considered one additional modification to the basic disk model discussed above, although it is not an alteration to the density structure. In Classical Be stars, there is some evidence that the $\alpha$ viscosity parameter required in hydrodynamical models of disk formation and dissipation is $\approx\,1$~\citep{Carciofi2012}. One possible interpretation of this result is that of sonic turbulence in the disk, i.e. $\nu=\alpha\,c_{S}\,H \sim c_{S}\,H$. For this reason, we have also considered models in which the disk is assumed to have a microtuburbulent velocity equal to the local sound speed. Microturbulence is a concept from classical stellar atmospheres that represents the dispersion of an assumed Gaussian distribution of turbulent velocities on scales smaller than unit optical depth. In this case, the turbulence acts to broaden the atomic absorption profile and hence is easily incorporated as an increase in the Doppler widths of radiative transitions. In our models, we assumed that the microturbulence value was either zero or equal to the local sound speed. These latter models will be referred to as {\it turbulent\/disks}.

The temperature structure and atomic level populations computed by {\sc Bedisk} are input into the code {\sc Beray}~\citep{Sigut2011}, which can compute observables such as line profiles, spectral energy distributions (hereafter SEDs) and monochromatic images in the sky. This is done by solving the equation of radiative transfer along a series of rays ($\approx10^5$) that pass through the star+disk system directed at the observer. Rays that terminate on the stellar surface use a Doppler-shifted photospheric (LTE) line profile for the initial boundary condition, while rays that pass entirely through the disk assume no incident radiation. Note that the {\sc Beray} calculation adds an additional parameter, namely the viewing inclination of the system ($i=0^{\circ}$ for a pole-on star face-on disk, and $i=90^{\circ}$ for an equator-on edge-on disk). Inclinations from $i=18^{\circ}$ to $i=75^{\circ}$ were computed. Finally, the computed spectral lines were convolved down to the instrumental resolution before comparing to the observed spectra.

\begin{table}[t]
\caption {Explored model parameters for disk of BD+65 1637.}
\label{tab:model_param}
\smallskip
\begin{center}{\small}
\resizebox{0.45\textwidth}{!}{
    \begin{tabular}{ll}
    \hline
    \hline
\noalign{\smallskip}
    Parameter & Range\\
\noalign{\smallskip}
\hline
\noalign{\smallskip}
Base Disk Density, $\rho_{0}$ $(\rm g\,cm^{-3})$ & $10^{-8}...10^{-13}$\\
\noalign{\smallskip}
Power Law Index, $n$ & $0.5...3.0$\\ 
\noalign{\smallskip}
Inclination, $i$ ($^{\circ}$) & $18...75$\\ 
\noalign{\smallskip}
Disk Radius, $R_{disk}$ ($R_{*}$) & $25...100$\\
    
\noalign{\smallskip}
\hline\
\end{tabular}
}
\end{center}
\end{table}

\subsection{Classical Be Stars and the {\sc Beray} and {\sc Bedisk} Codes}
\label{sec:classicalbestars}

{\sc Bedisk} and {\sc Beray} are non-LTE radiative transfer codes constructed specifically for Classical Be stars and their ionized, gaseous decretion disks. Many studies, such as~\cite{Silaj2014} and~\cite{Silaj2010}, have been able to successfully model the structure of the gaseous disk by comparing synthetic and observed H$\alpha$ lines. ~\cite{Sigut2011} has shown that such models based on H$\alpha$ are also able to correctly predict observed Fe\,{\sc ii} lines in the spectra of these stars. These models have also been able to reproduce the observed correlation seen between H$\alpha$ and long term variations in visual magnitude, which are interpreted as formation and dissipation of the disk over a long periods of time in Classical Be stars~\citep{SP2013}. Observed IR line fluxes~\citep{Jones2009,Halonen2008} as well as optical and near-IR interferometry~\citep{Jones2008,Tycner2008,Mackay2009,Grzenia2013,Sigut2015} computed with {\sc Bedisk} and {\sc Beray} models, have been used to put constraints on the several Classical Be star disks.

Several authors, such as~\cite{Carciofi2009,Carciofi2007,Carciofi2006}, studied the Classical Be stars $\alpha$ Eri, $\zeta$ Tau and $\delta$ Sco by fitting viscous decretion disk models to the observed Balmer lines, SEDs, and polarization measurements.~\cite{Silaj2010} studied 56 Be stars and successfully fit the observed H$\alpha$ profiles to {\sc Bedisk} models. Many individual stars, such as $\chi$ Oph~\citep{Tycner2008}, $\kappa$ Dra, $\beta$ Psc, $\upsilon$ Cyg~\citep{Jones2008}, and \textit{o} Aqr~\citep{Sigut2015} have been studied spectroscopically as well as interferometrically, and are found to match a density model similar to the one adopted in this study. All of these mentioned studies, as well as several others, have fit the observed line profiles well with power law index ($n$) ranging from 2 to 4, typically 3.5, and the disk base density varying between $10^{-10}$ and $10^{-12}$ $~\rm g\,cm^{-3}$ ~\citep{Rivinius2013}.

Given the noted similarities between HBe and Classical Be stars~\citep{HP1992,BB2000,Mottram2007}, a good starting point for the modeling of the emission spectra of HBe stars is using codes that have successfully been able to reproduce emission lines from the gaseous disks of Classical Be stars. The analysis can give insights on the regions where the lines are being formed, the mass of the disk, and the temperature and density structure of the emitting regions.

\section{Results}
\label{sec:results}

Large libraries of synthetic line profiles were calculated for H$\alpha$,
H$\beta$, the Ca~{\sc ii} IR triplet ($\lambda\,8542$), Fe\,{\sc
ii} ($\lambda\,5169$,$\lambda\,5317$)\footnote{For Fe\,{\sc ii} 5317, only the multiplet 49 component was included.}, and He~{\sc i} ($\lambda\,6678$)
for disks surrounding a B2 star using combinations of
the disk density parameters listed in Table~\ref{tab:model_param}.
Values of the disk base density parameter, $\rho_{0}$, ranged from
$10^{-13}$ to $10^{-8}\,\rm gm\,cm^{-3}$, and the power-law index $n$
ranged from $0.5$ to $3.0$. Three different sized disks
were considered, R$_{disk}$=$25$, $50$ and $100$\,R$_{*}$; thus, the largest
disk considered has an outer diameter of $3.1\,$AU. This
range of disk density parameters and disk sizes includes the range of
values typically found for Classical Be stars, as noted above, but with an extension to
more massive disks (i.\ e.\ higher $\rho_0$ and/or lower $n$). All synthetic line profiles were calculated at viewing inclinations of 18$^{\circ}$,
45$^{\circ}$, 60$^{\circ}$ and 75$^{\circ}$, which represent the centers of the first four bins of five equal-area bins in a random $\sin i$ distribution.

Each observed line profile was compared to its synthetic library by computing a figure-of-merit, ${\cal F}$, defined as
\begin{equation}
{\cal F} \equiv \frac{1}{N}\,\sum_{i=1}^{N}\,\frac{|F_i^{\rm Mod}-F_{i}^{\rm Obs}|}{F_{i}^{\rm Obs}}, \,
\label{eq:FOM}
\end{equation}
where $F_{i}^{\rm Obs}$ is the observed relative flux, $F_i^{\rm Mod}$
is the model relative flux, and the sum is over the $N$ wavelength
points spanning the line. In performing this sum, a range of small shifts
to the observed wavelength scale was also tried, within the errors
of the star's radial velocity. The smallest value of ${\cal F}$ was
deemed to define the best-fit model for that feature, although all profiles with small values of ${\cal F}$ were visually inspected. In addition, the disk density parameters of profiles that fit the observed
profiles almost as well as the best-fit model were also examined, and
this point, concerning the uniqueness of the fits ,will be discussed in Section~\ref{sec:uniquediskmodels}.

While the minimum of ${\cal F}$ for a given line, say $\cal F^{\rm
H\alpha}$, defines the best fit for that particular line, it
is not guaranteed that the best fit model for all lines will result in
the same set of disk density  parameters. A figure-of-merit defined
by Eqn.~\ref{eq:FOM} can be obtained for each line considered, i.e.\
$\cal F^{\rm H\alpha}$, $\cal F^{\rm H\beta}$, etc...\ Therefore, it is
possible to search for the best set of disk parameters that minimizes
the sum of all of the line figures-of-merit i.e. the global, best-fit model. 

For all the line profile matches performed in this study, the effort was made to fit to the blue peak of the emission line. When a reasonable fit was not found for the blue peak, the fit was computed for the red peak instead. 

We will now first discuss the best-fit models for each line individually, and then consider the best global model.

\subsection{Individual Fits}
\label{sec:linebestfits}

\begin{figure*}[h]
\centering
\epsscale{2.2}
\plotone{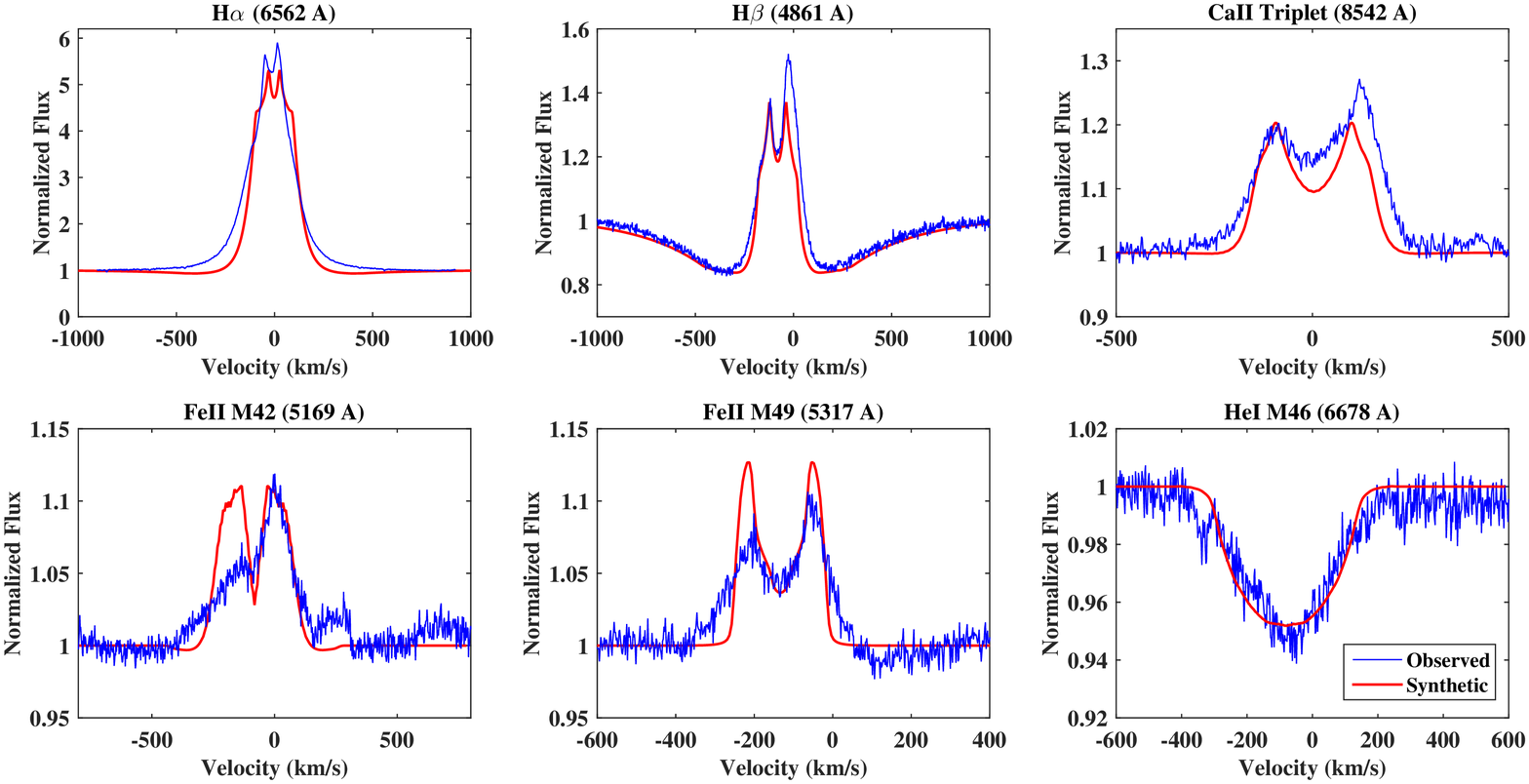}
\caption{The best-fit for each line synthetic line profile (red) for the observed emission line (blue) of H$\alpha$ ($\lambda\,6562$), H$\beta$ ($\lambda\,4861$), Ca\,{\sc ii} IR Triplet ($\lambda\,8542$), Fe\,{\sc ii} multiplet (42) ($\lambda\,5169$) and Fe\,{\sc ii} multiplet (49) ($\lambda\,5317$) and He\,{\sc i} ($\lambda\,6678$) for BD+65\,1637, which were modeled using {\sc Beray}. The fit parameters for each model can be found in Table~\ref{tab:bestfitmodels}, with the exception of He\,{\sc i} as it is fit by a photospheric profile.}
\label{fig:bd_bestmatches_alllines}
\end{figure*}

The best-fit models for all individual lines are listed
in Table~\ref{tab:bestfitmodels}, and the best synthetic line profile
fits to the individual observed emission lines are shown in
Figure~\ref{fig:bd_bestmatches_alllines}. With the freedom to chose the density model independently for each line, the observed line profiles
can be reproduced quite well in strength, shape, and equivalent width
by the models.

\begin{table*}[t]
\caption{Best-fit model parameters for individual emission lines and the
global models with and without Ca\,{\sc ii}.}
\label{tab:bestfitmodels}
\smallskip
\begin{center}{\small
\resizebox{0.99\textwidth}{!}{
    \begin{tabular}{cccccc}
    \hline
\hline
\noalign{\smallskip} \\
    Emission Line & Disk Density $\rho_{0}$ $(\rm g\,cm^{-3})$ & Power Law Index $n$ & Inclination $i$ ($^{\circ}$) & Disk Radius $R_{disk}$ ($R_{*}$) & Model Type\\ 
\noalign{\smallskip}
\hline \\
\noalign{\smallskip}
    H$\alpha$ (6562 \AA) & 3.2\,*10$^{-12}$ & 2.0 & 60 & 50 & Thin\\ \\
\noalign{\smallskip}
%\tableline
\noalign{\smallskip}
    H$\beta$ (4861 \AA) & 1.0\,*10$^{-11}$ & 2.0 & 45 & 25 & Thin\\ \\
\noalign{\smallskip}
%\tableline
\noalign{\smallskip}
    Ca\,{\sc ii} IR-triplet (8542 \AA) & 1.0\,*10$^{-10}$  & 2.0 & 60 & 25 & Thin \& Turbulent \\ \\
\noalign{\smallskip}
%\noalign{\smallskip}
  Fe\,{\sc ii} (5169 \AA) & 1.0\,*10$^{-10}$ & 3.0 & 45 & 25 & Thick \& Turbulent\\ \\
\noalign{\smallskip}
  Fe\,{\sc ii} (5317 \AA) & 1.0\,*10$^{-9}$ & 1.5 & 75 & 25 & Thin \& Turbulent\\ \\
 %HeI (6678 \AA) & 1.0*10$^{-12}$ & 2.0 & 45 & 2 & Thin  \\
 \hline \\  \\
 Global & 1.0\,*10$^{-10}$ & 2.0 & 45 & 50 & Thin  \\ \\

 Global (w/o Ca\,{\sc ii}) & 1.0\,*10$^{-10}$ & 3.0 & 45 & 50 & Thin \& Turbulent  \\ \\
 
\noalign{\smallskip}
\hline \\	 
\end{tabular}}}
\end{center}
\tablecomments{No best-fit model for He\,{\sc i}($\lambda\,6678$) absorption line is given, as it is of photospheric origin.}
\end{table*}

The best fit for H$\alpha$ is a 50\,R$_{*}$, thin
disk model with disk density parameter $\rho_{0}$ of 3.2\,*10$^{-12}\,\rm
g\,cm^{-3}$ and power law index $n$ of 2.0 seen at at an inclination
of 60$^{\circ}$. The observed and synthetic profiles are compared in
Figure~\ref{fig:bd_bestmatches_alllines}. The width of H$\alpha$
at its base is underestimated, and a better fit might be possible by refining
the viewing inclination, however, we have not attempted this.

For H$\beta$, the best match to the observed profile was found for a
model with slightly smaller, thin disk of 25\,R$_{*}$ seen at 45$^{\circ}$
with disk density parameter $\rho_{0}$ of $1.0\,*10^{-11}\,\rm g\,cm^{-3}$
and power law index $n$ of 2. The overall strength and width of
H$\beta$ (including its absorption wings) are well reproduced by the
model. We note that the $V/R$ asymmetry cannot be reproduced by our assumed
axisymmetric disk models.

At this point, we immediately see that the best-fit models for H$\alpha$
and H$\beta$ differ. Nevertheless, it should be kept in mind that
in addition to the best-fit model, there is a range of other disk models
that fit each profile nearly as well. For example, there will be {\bf N} models
that fit H$\alpha$ with a figure of merit within 25\% of the best fit model, and for H$\beta$, there will be {\bf M} such models. We will return to the question of the number of such models and how the disk density parameter ranges compare in Section~\ref{sec:uniquediskmodels}.

For Ca\,{\sc ii} $\lambda\,8542$, the best-fit model to the observed
line has an disk density parameter $\rho_{0}$ of 1.0\,*10$^{-10}$ $\rm
g\,cm^{-3}$ and power law index $n$ of 2 seen at 60$^{\circ}$ for
a 25\,R$_{*}$ thin and turbulent disk. We again note that the width and overall strength of the line are well reproduced.

For the two Fe\,{\sc ii} lines, the figure of merit ${\cal F}$ was computed by using only the red half of the line, i.e.\ the blue peak was ignored in the fit.
The Fe\,{\sc ii} multiplet (42) $\lambda\,5169$ line requires a disk density parameter $\rho_{0}$ of 1.0\,*10$^{-10}$ $\rm g\,cm^{-3}$ and a power law index $n$ of 3 seen at 45$^{\circ}$ for a 25\,R$_{*}$ thick and turbulent
disk. The Fe\,{\sc ii} multiplet (49) $\lambda\,5317$ line requires a model with disk density parameter $\rho_{0}$ of 1.0\,*10$^{-9}$
$\rm g\,cm^{-3}$ and a power law index $n$ of 1.5 seen at 75$^{\circ}$
for a 25\,R$_{*}$ thick and turbulent disk. We note that the lines of Ca\,{\sc ii} and Fe\,{\sc ii} prefer the turbulent disk model, as these models tend to produce
broader and stronger lines.

\begin{figure*}[h]
\centering
\epsscale{2.2}
\plotone{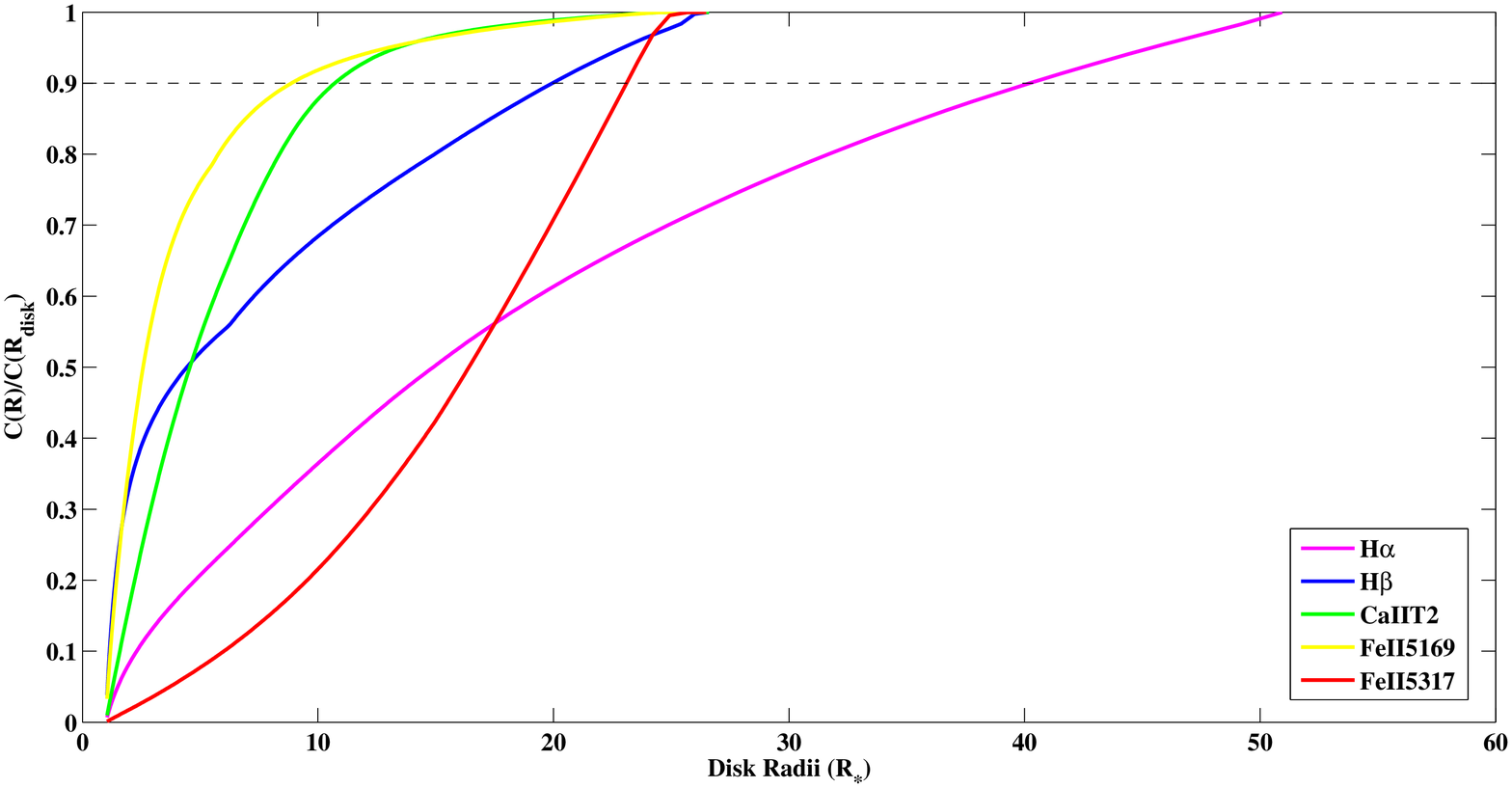}
\caption{The cumulative line emission (Equation~\ref{eqn:c}) as a function of distance from the star for each individual line using the best-fit disk density model. The model parameters that produced each individual line can be found in Table~\ref{tab:bestfitmodels}. The black dashed line represents a cumulative emission fraction of $0.9$.}
\label{fig:bd_bestlinematches_cumfluxvsr}
\end{figure*}

To investigate what range of disk radii contribute to the formation
of lines considered, the cumulative intensity produced by each emission
line was plotted against the radius of the disk for the models listed in Table~\ref{tab:bestfitmodels} as shown in Figure~\ref{fig:bd_bestlinematches_cumfluxvsr}. To do this, a face-on synthetic image ($i=0^{\circ}$) was produced using the best-fit disk density model for each line. For each $i=0^{\circ}$ image, the intensity was integrated over the total width of the line. Then, the integrated intensity out to a disk radius of $R$ can then be defined as

\begin{equation}
C(R)=2\pi\,\int_{R_*}^{R} I(R^{\prime})\,R^{\prime}\,dR^{\prime} \,,
\label{eqn:c}
\end{equation}

where $I(R)$ is the wavelength integrated line intensity at distance $R$, and $R_{*}$ is the stellar radius, assumed to be the inner edge of the disk. Then $C(R)/C(R_{disk}$) can be plotted versus $R$ to determine how the line intensity is accumulated by the disk.
In the Figure~\ref{fig:bd_bestlinematches_cumfluxvsr}, a solid
black line shows the cumulative fraction of $0.9$.
It is important to keep in mind when looking at this figure that the disk density
model particular to each transition has been used and not a single disk density
model. This explains, for example, why C=1 is reached at 50\,R$_{*}$ for H$\alpha$, but 25\,R$_{*}$ for the remaining lines. In order to
reproduce the strength of the H$\alpha$ emission, an extended emission region
is required, reaching 90\% of the emission at 40\,R$_{*}$. However, 90\% of the emission for Ca\,{\sc ii}, and Fe\,{\sc ii} ($\lambda\,5169$) originates from the inner most 10\,R$_{*}$ of the disk, and H$\beta$ and Fe\,{\sc ii} ($\lambda\,5317$) are intermediately reaching 90\% complete at $\sim$20\,R$_{*}$.

This figure also illustrates how the disk might be structured in order
to produce all the line profiles by having the disk's equatorial density vary in a more general way than as a single power-law (see Section~\ref{sec:discussion}).

\subsection{Global Fits}
\label{sec:globalbestfits}

Given that different disk density models are required to best-fit each observed profile for BD+65\,1637, the next logical step was to see if a single disk density model could fit all the lines in a reasonable (as opposed to optimal) manner. This will also assist us in deciding how to move forward in looking for a more general density model that would better describe the structure of the disk. 
To find the single best model, we minimized the sum of all the individual ${\cal F}$, i.e. 
\begin{equation}
{\cal F^{\rm total}} = \sum_{i=1}^{6}\, w_{i} {F^{\rm i}} \,
\label{eq:FOM_total}
\end{equation}
where $i$ ranges over the six lines considered. Initially, we set $w_{i}$=1 for all $i$ to weigh all six lines equally. The model that was found to best reproduce all the observed line profiles in this manner is listed in Table~\ref{tab:bestfitmodels}.

\begin{figure*}[h]
\centering
\epsscale{2.2}
\plotone{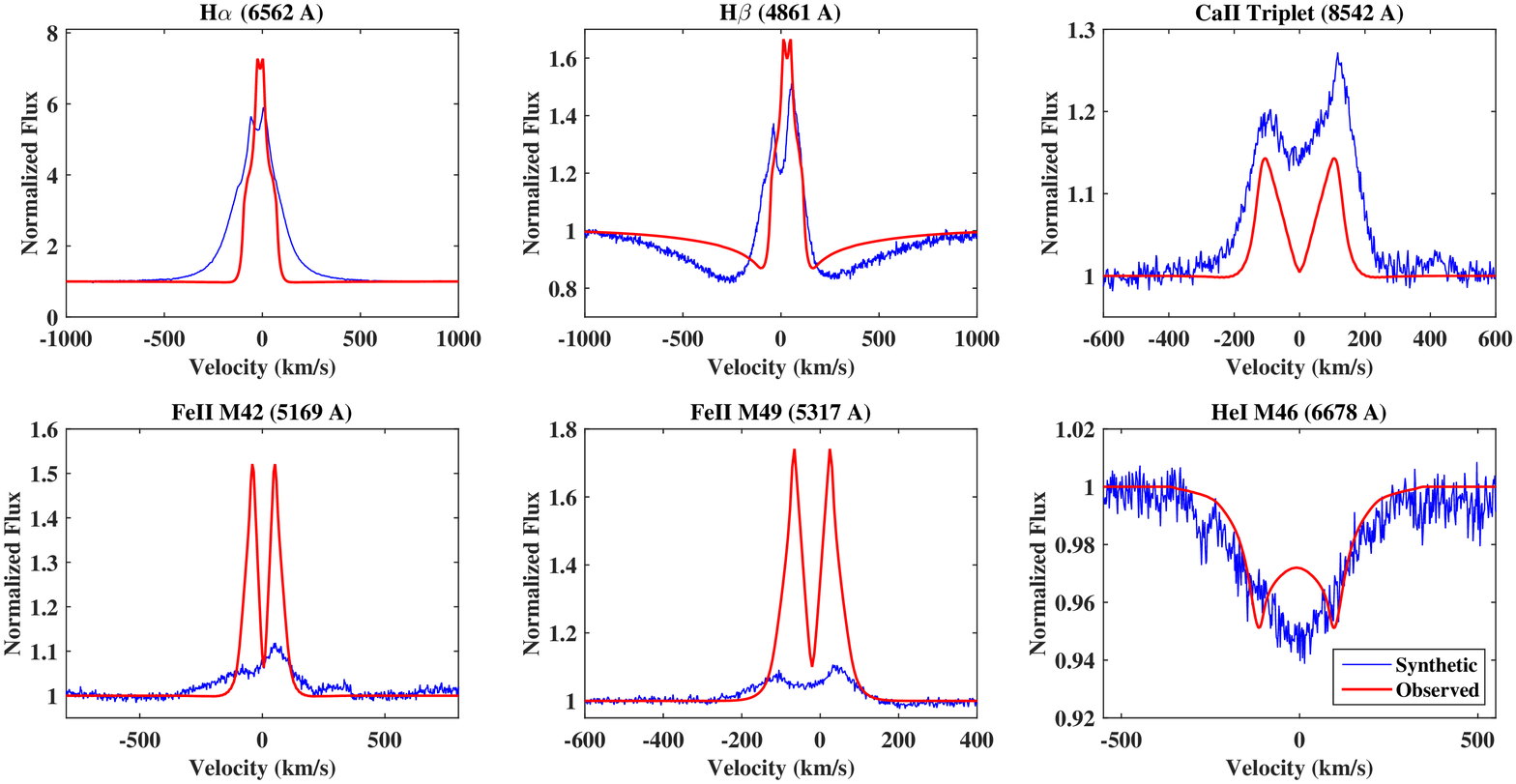}
\caption{The global best-fit for all the synthetic line profiles (red) for the observed emission lines (blue) of H$\alpha$ ($\lambda\,6562$), H$\beta$ ($\lambda\,4861$), Ca\,{\sc ii} IR Triplet ($\lambda\,8542$), Fe\,{\sc ii} multiplet (42) ($\lambda\,5169$) and Fe\,{\sc ii} multiplet (49) ($\lambda\,5317$) and He\,{\sc i} ($\lambda\,6678$) for BD+65\,1637, which were modeled using {\sc Beray}.  The global best fit model parameters used for this Figure can be found in Table~\ref{tab:bestfitmodels}.}
\label{fig:bd_bestmatches_global}
\end{figure*}

Figure~\ref{fig:bd_bestmatches_global} shows all six synthetic emission lines produced for this model as compared to the observed line profiles. The density parameters are a power law index $n$ of 2, a disk density parameter $\rho_{0}$ of 1.0\,*10$^{-10}$ $\rm g\,cm^{-3}$ with a 50\,R$_{*}$, thin disk seen at 45$^{\circ}$. As illustrated in the figure, the Balmer lines can be reproduced approximately in strength, but are too narrow at the base; the metal lines are either too strong (Fe\,{\sc ii} lines) or too weak (Ca\,{\sc ii} IR Triplet) compared to the observed emission lines. The mismatch of the shape of the Balmer line profiles, particularly in the wings, indicates that that the material is not distributed correctly in the disk by a single power-law. The synthetic line profile for Ca\,{\sc ii} IR triplet $\lambda\,8542$ is weaker in strength, as well as narrower in velocity, than the observed line profile. It does, however, produce a double-peaked shape. Both of the synthetic Fe\,{\sc ii} line profiles have approximately the same shape and strength when compared to each other; however the width of the wings are different; Fe\,{\sc ii} (42) $\lambda\,5169$ has narrower spread of velocity in the wings when compared to Fe\,{\sc ii} (49) $\lambda\,5317$. Both synthetic Fe\,{\sc ii} line profiles are too strong compared to the observed profiles, suggesting that smaller regions may be required to reproduce the observed line (as seen in Section~\ref{sec:linebestfits}). Finally, He\,{\sc i} shows absorption with some central emission, sometimes called a central quasi emission (CQE) feature~\citep{Hanuschik1995}. As the star and disk system is seen 45$^{\circ}$, the CQE can be attributed to the disk which partly blocks the direct stellar radiation.

\begin{figure*}[h]
\centering
\epsscale{2.2}
\plotone{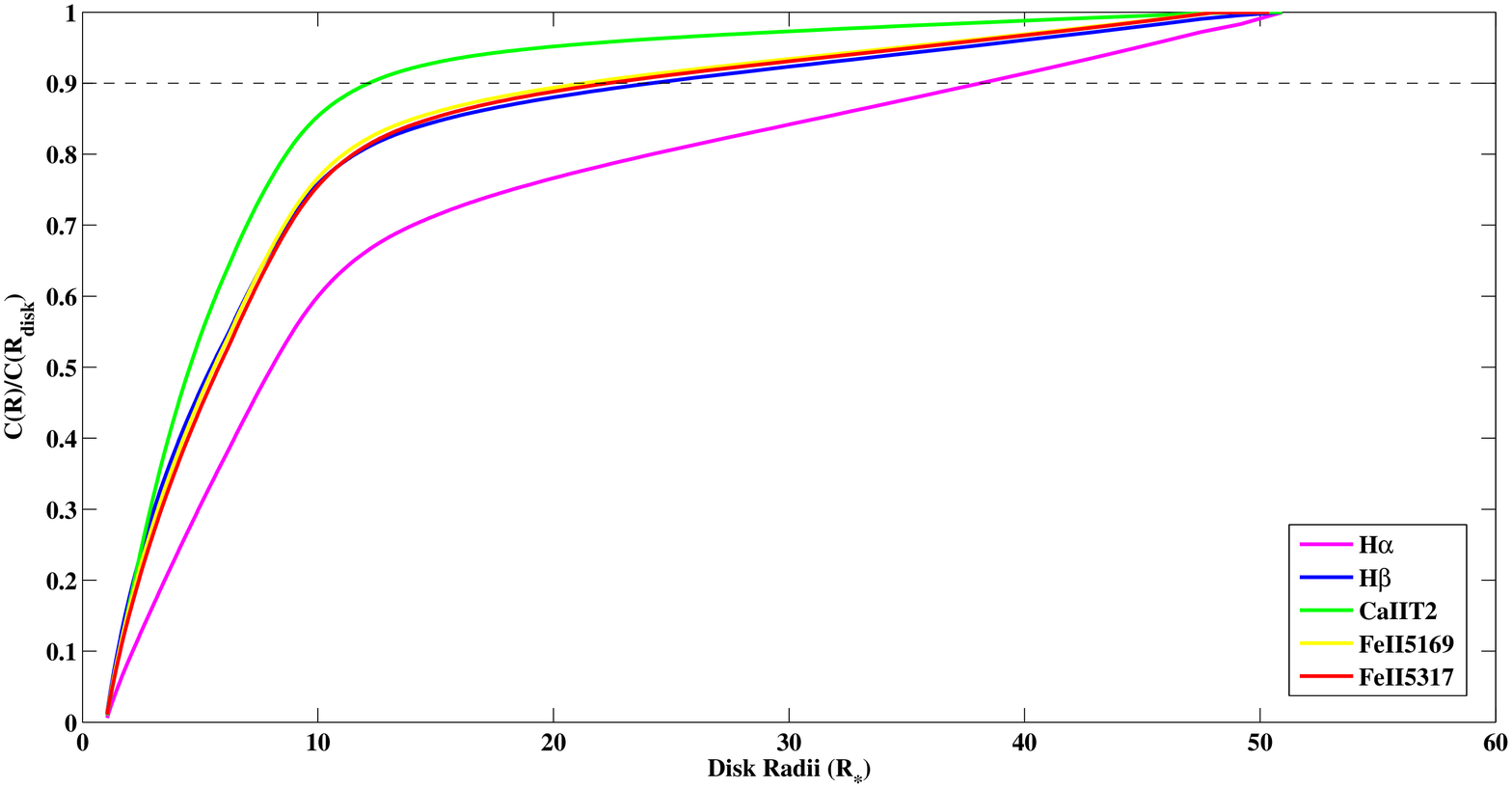}
\caption{The cumulative line emission (Equation~\ref{eqn:c}) as a function of distance from the star for each individual line using the global disk density model. The global model parameters can be found in Table~\ref{tab:bestfitmodels}. The black dashed line represents from where the 90\% of the emission is coming from.}
\label{fig:bd_bestglobalmatches_cumfluxvsr}
\end{figure*}

Figure~\ref{fig:bd_bestglobalmatches_cumfluxvsr} illustrates where the intensity is produced by these lines for this single power law model by plotting C(R), Equation~\ref{eqn:c}, as a function of disk radius. It can be seen that in all the cases, 90\% of the emission now is coming from inside 40\,R$_{*}$ (or 1.25 AU). This figure also illustrates that the Ca\,{\sc ii} IR Triplet ($\lambda\,8542$) is produced in the innermost 10\,R$_{*}$ of the disk; the H$\beta$ ($\lambda\,4861$) and the Fe\,{\sc ii} multiplets ($\lambda\,5169$ \& $\lambda\,5317$) produce most of their emission within 20\,R$_{*}$, and H$\alpha$ emission is produced throughout the disk, with 90\% coming from within 40\,R$_{*}$. Comparison of Figures~\ref{fig:bd_bestlinematches_cumfluxvsr} and~\ref{fig:bd_bestglobalmatches_cumfluxvsr} shows that the Ca\,{\sc ii} IR triplet forms in the innermost part of the disk, while H$\alpha$ forms throughout the disk. The emission from H$\beta$ and Fe\,{\sc ii} multiplets are intermediate and emerge from the same region for the global model. 

\begin{figure*}
\centering

\begin{subfigure}[t]{0.8\textwidth}
%\epsscale{1.5}
%\plotone{Figure11.eps}
\includegraphics[width=0.98\textwidth]{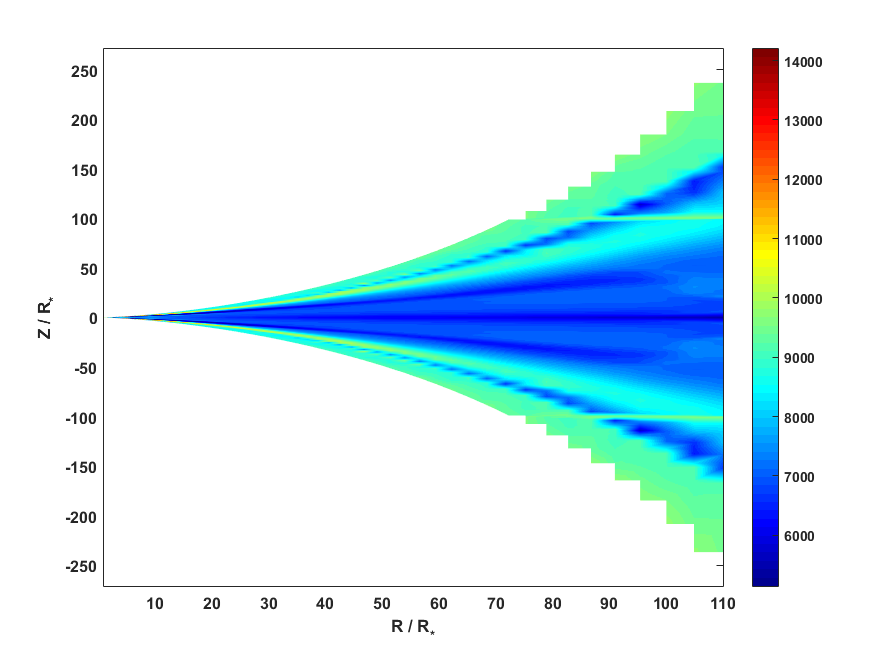}
\label{fig:bd_temperaturezoomedout}
\end{subfigure} %delete end{subfigure} from here if using plotone & eps command

\quad
\begin{subfigure}[t]{0.8\textwidth}
%\epsscale{1.5}
%\plotone{Figure12.eps}
\includegraphics[width=0.98\textwidth]{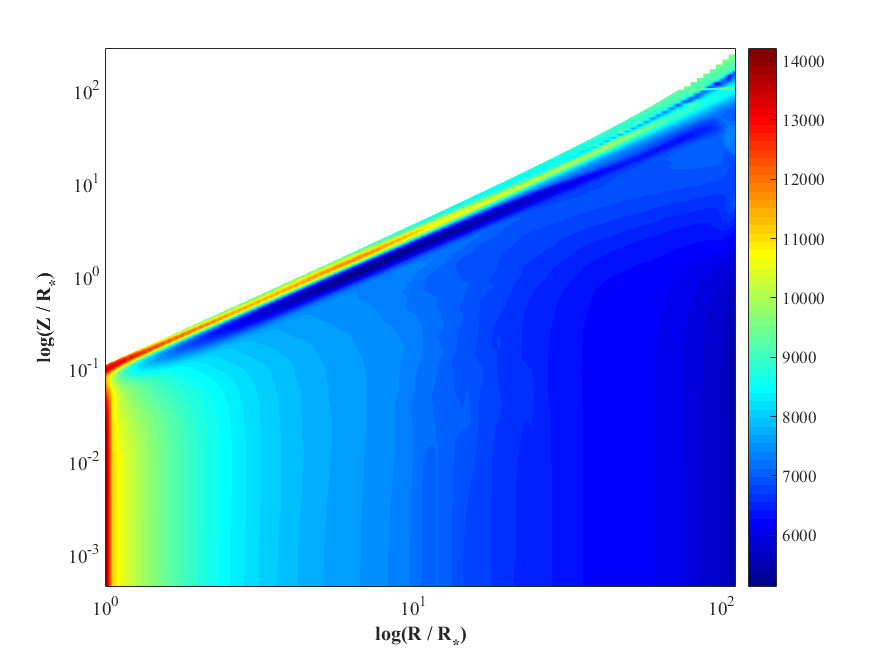}
\label{fig:bd_temperature_zoomedin_loglog}
\end{subfigure}
\caption{Temperature distribution in the disk for a model with disk density parameter $\rho_{0}$ of $1.0*10^{-10}~\rm g\,cm^{-3}$ and power law index $n$ of 2.  On the top, the plot illustrates the temperature structure in the entire disk, while at the bottom, in a log-log plot, the temperature distribution in the region close to the stellar surface is illustrated. A colorbar giving the temperature scale (in K) is to the right of each plot.}
\label{fig:bd_temperature}
\end{figure*}

In addition to constraining the density distribution in the disk for a model that is consistent with the observations, the global model can be used to give insight to the temperature structure of the disk. Figure~\ref{fig:bd_temperature} illustrates the temperature distribution predicted by {\sc Bedisk} in a thin disk for a model with a disk density parameter $\rho_{0}$ of $1.0\,*10^{-10}~\rm g\,cm^{-3}$ and power law index $n$ of 2. The upper plot illustrates the temperature in the entire disk, which generally ranges from 5500 K to 10000 K. The bottom log-log plot shows the region close to the stellar surface where temperatures in the disk can reach as high as 14000 K. When combined with the density structure, this provides valuable information on structure of the inner, gaseous disk. For example, even at 110\,R$_{*}$, the coolest temperature predicted in the equatorial plane, $\simeq$5500 K, is still above the dust sublimation temperature.

Finally, the disk density parameters can be used to estimate the total mass of the inner gaseous disk. The mass is estimated to be $9.3\,*10^{26}~\rm gm$ ($5.7\,*10^{-8}\,M_{*}$~or $4.6\,*10^{-7}$ M$_{\sun}$) while the scale height $H$ of the disk at the stellar surface was estimated to be $1.6\,*10^{10}~\rm cm$ ($3.5\,*10^{-2}\,R_{*}$~or $0.23$ R$_{\sun}$).

\begin{figure*}[h]
\centering
\epsscale{2.2}
\plotone{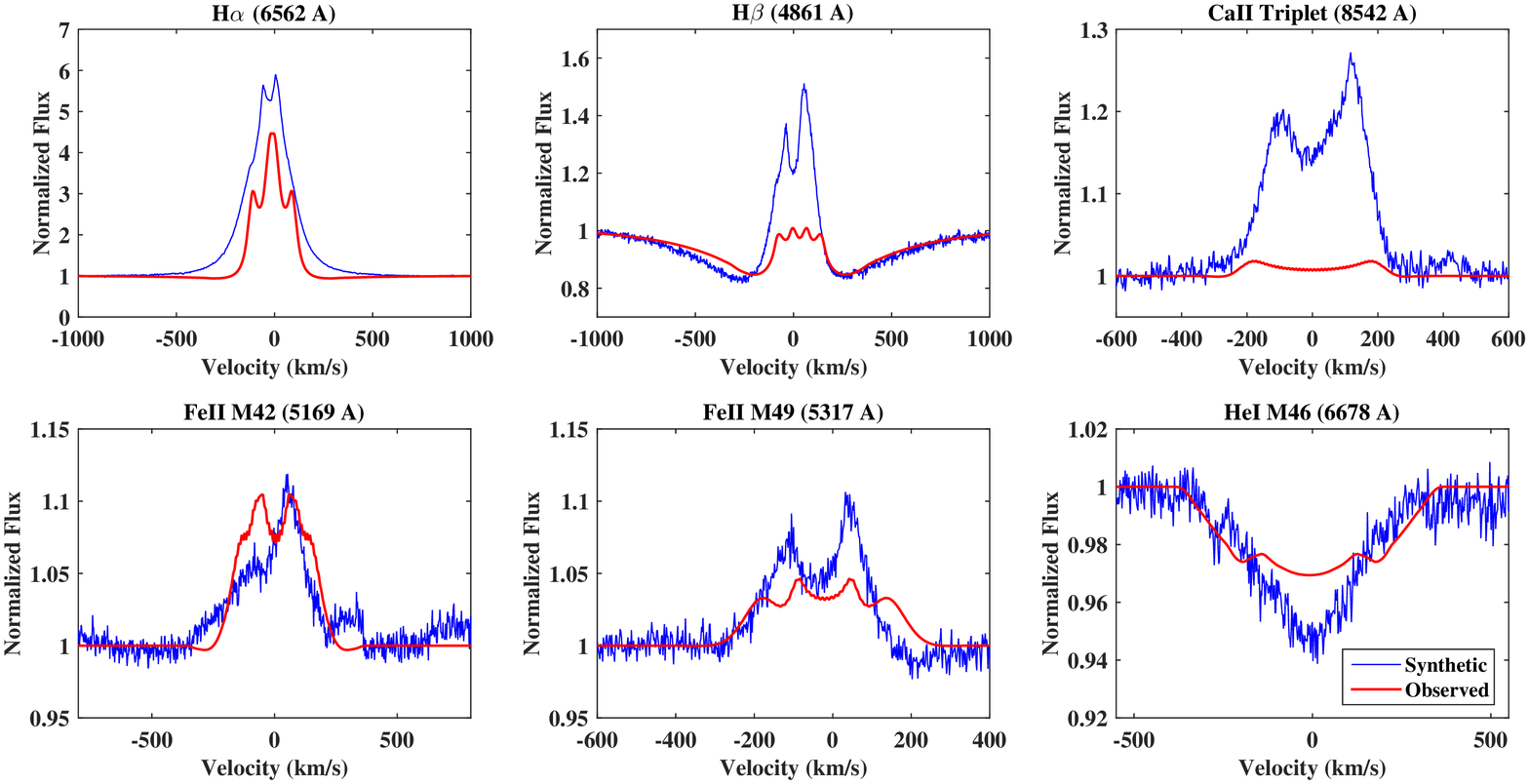}
\caption{The global best-fit for the model that generates reasonable emission for all lines except Ca\,{\sc ii} IR Triplet ($\lambda\,8542$) when compared to the observed spectral lines (blue). The global model used can be found in Table~\ref{tab:bestfitmodels} (on the line "w/o Ca\,{\sc ii}").}
\label{fig:bd_globalfits_alllinesexceptCaII}
\end{figure*}

It was noticed in the fitting process that the metal lines, especially the Ca\,{\sc ii} IR Triplet, require a high disk density parameter $\rho_{0}$ with low value of the power law index $n$.~Ca\,{\sc ii} is generally not well reproduced by the models which otherwise are found to work reasonably well for a single power law. For this reason, we searched for global fits excluding the Ca\,{\sc ii} line by setting $w_{i}=0$ for $\cal F^{\rm Ca\,II}$ in Equation~\ref{eq:FOM_total}. The result can be seen in Figure~\ref{fig:bd_globalfits_alllinesexceptCaII}. The details for this model can be found in Table~\ref{tab:bestfitmodels}, and this model is able to reproduce the emission in the two Fe\,{\sc ii} lines reasonably well. The Balmer lines are not strong enough to match the strength of the observed profile. However they match well when the width of the lines is considered. The same is the case for He\,{\sc i} line. The Ca\,{\sc ii} IR triplet for this model shows hardly any emission, and this indicates that Ca\,{\sc ii} is likely formed in a different region, while all the other lines can be reasonably produced by disk with a single power law with $n=3$. The mass of this disk was estimated to be $4.8\,*10^{25}~\rm gm$ ($2.3\,*10^{-9}\,M_{*}$ or $4.6\,*10^{-7}$ M$_{\sun}$).

\begin{figure*}[h]
\centering
\epsscale{2.2}
\plotone{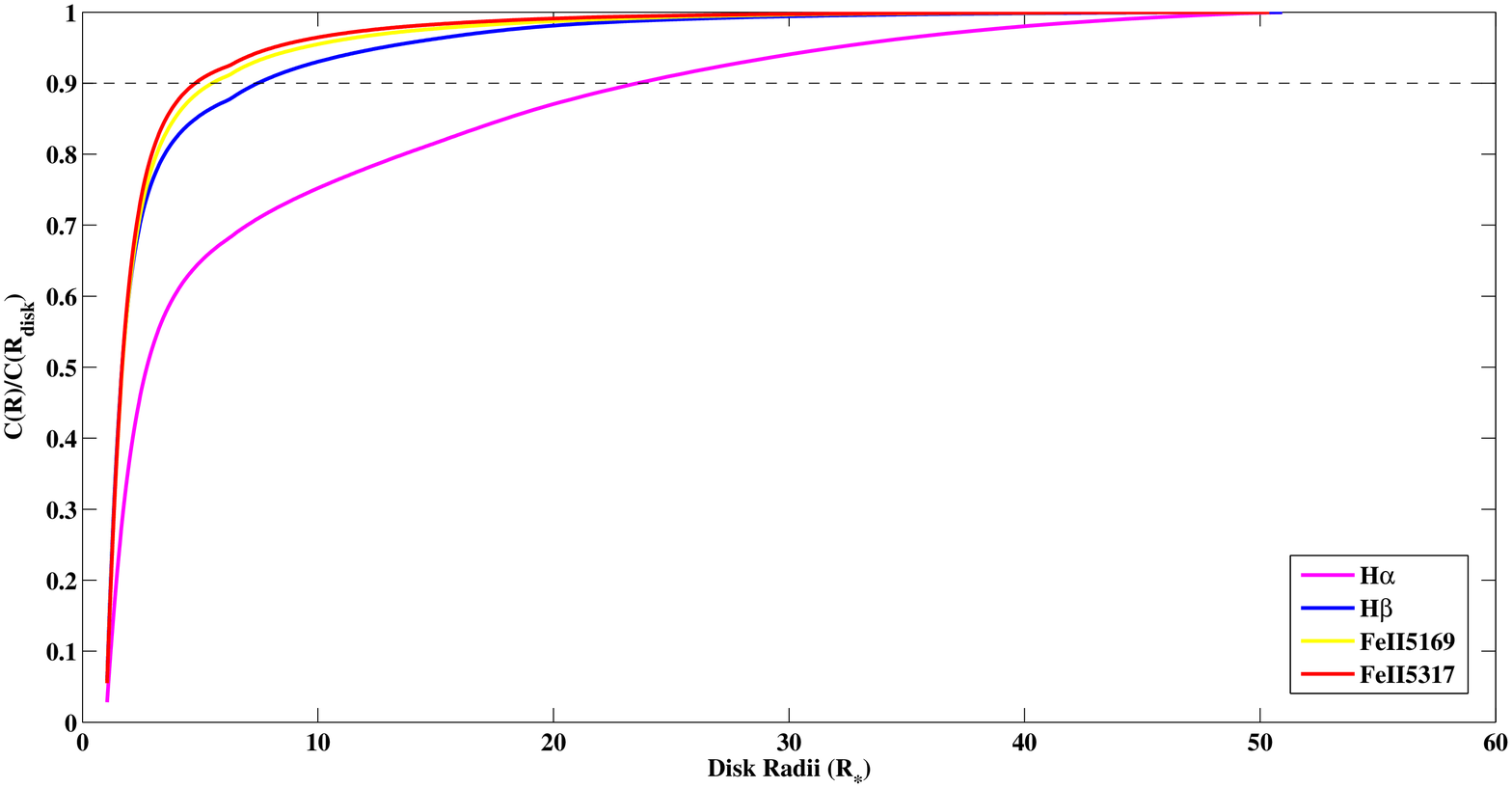}
\caption{The cumulative line emission (Equation~\ref{eqn:c}) as a function of distance from the star for each individual line using the global disk density model without Ca\,{\sc ii}. The model parameters used for this figure can be found in Table~\ref{tab:bestfitmodels}. The black dashed line represents from where the 90\% of the emission is coming from.}
\label{fig:bd_globalfits_alllinesexceptCaII_cumfluxvsr}
\end{figure*}

In order to illustrate where these lines are formed in this model which excludes Ca\,{\sc ii}, and if there is any similarity to the previous global model, Figure~\ref{fig:bd_globalfits_alllinesexceptCaII_cumfluxvsr} was constructed. The plot shows H$\alpha$ forming almost throughout the entire disk with the 90\% of emission coming from inside the 30\,R$_{*}$. H$\beta$, Fe\,{\sc ii} ($\lambda\,5169$) and Fe\,{\sc ii} ($\lambda\,5317$) can be seen forming within 15\,R$_{*}$. When compared to the previous global fit model (Figure~\ref{fig:bd_bestglobalmatches_cumfluxvsr}), all the emission lines except H$\alpha$ in this model are produced within half the radius. 

Thus, from the all three models considered, it can be concluded that the H$\beta$ and the metal lines form in the innermost region of the disk while the H$\alpha$ forms in an extended region covering nearly the entire disk.

\subsection{The Near-IR SED}
\label{sec:sed}
\begin{figure*}
\centering
\epsscale{1.35}
\plotone{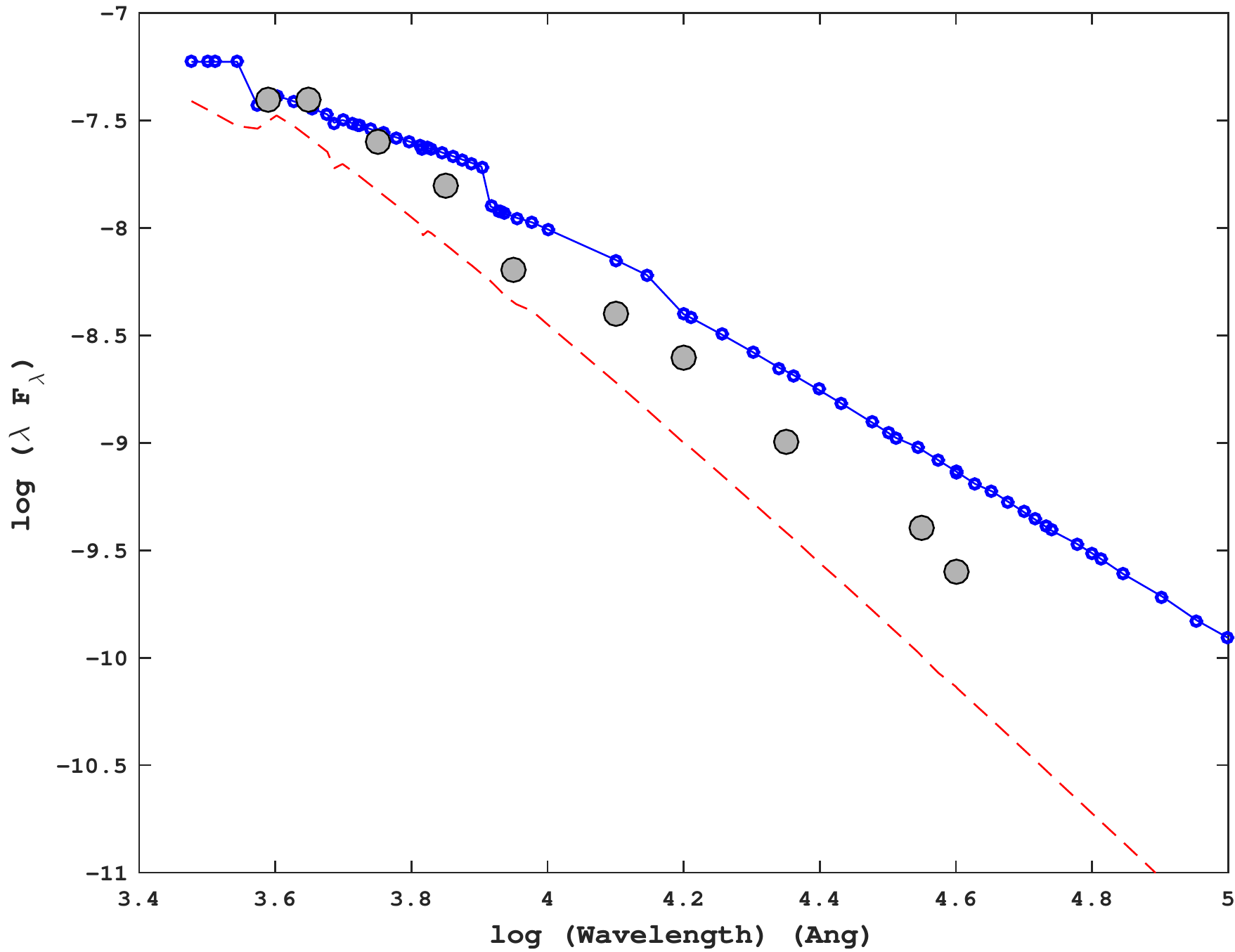}
\caption{The continuum SED calculated using {\sc Beray} for the global best fit model (with Ca\,{\sc ii}) for the disk (blue circles).  The photospheric spectrum of the star is shown as the red dashed line. These SED observations of~\cite{Hillenbrand1992} are marked with gray filled circles. The global model used for this SED of the disk can be found in Table~\ref{tab:bestfitmodels}.}
\label{fig:bd_sed}
\end{figure*}

As mentioned in Section~\ref{sec:modeling}, {\sc Beray} can also calculate continuum SEDs of the star+disk system. In order to assess how comparable these models are to the available observations, a SED was produced for the global disk model of Table~\ref{tab:bestfitmodels} and compared to the observed SED for BD+65\,1637 found in~\cite{Hillenbrand1992}. This is illustrated in Figure~\ref{fig:bd_sed}. The star's continuum SED, i.e. in the absence of a disk, is also shown. As it can be seen in the figure, the global disk model produces a brighter SED at longer wavelengths compared to the observed SED. This suggests a thinner and less dense disk than those considered here is required in order to be comparable to the observed SED. However, it is important to note that the SED observations were taken more than 16 years prior to the observations of the emission lines used in the analysis here. H$\alpha$ has been previously reported to be variable in EW (-45~\AA~to -26~\AA) and hence, the comparison of the disks over a long period of time should be considered with caution. 

\section{Uniqueness of Disk Models}
\label{sec:uniquediskmodels}

\begin{figure*}[h]
\centering
\epsscale{1.35}
\plotone{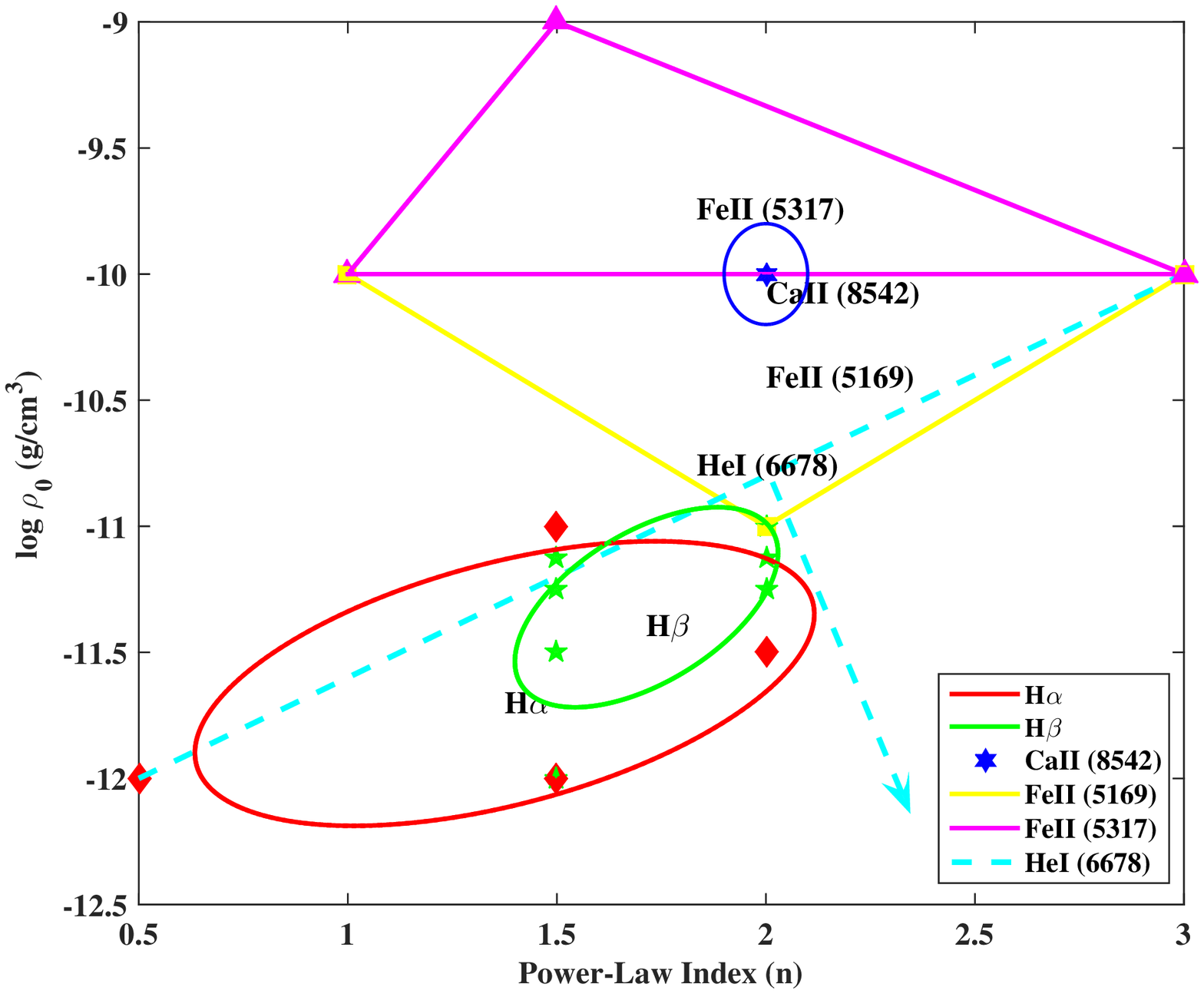}
\caption{Regions of the (n, log\,$\rho_{0}$) plane occupied by the top 25\% of the profiles that match the observed profile individually (using the figure of merit values ${\cal F}$). H$\alpha$ ($\lambda\,6562$) in shown in Red, H$\beta$ ($\lambda\,4861$) in Green, Ca\,{\sc ii} ($\lambda\,8542$) in Blue, Fe\,{\sc ii} multiplet (42) ($\lambda\,5169$) in Yellow and multiplet (49) ($\lambda\,5317$) in Magenta, and He\,{\sc i} ($\lambda\,6678$) in Cyan. the He\,{\sc i} constraint is the region that results in no observable disk emission in the He\,{\sc i} photospheric line.}
\label{fig:bd_FOMEllipse}
\end{figure*}

As described in the Section~\ref{sec:results}, the fitting procedure used the values of the figure of merit ${\cal F}$ for all the lines, found by using Equation~\ref{eq:FOM_total}, to build a set of the global, best-fit models. In Section~\ref{sec:linebestfits}, it was noted that although one model is the best-fit for each line profile, more than one model can fit a particular line profile within a certain range of ${\cal F}$. Table~\ref{tab:25FOM_models} gives the number of models for each emission line that have ${\cal F}<1.25{\cal F}_{min}$ (i.e. the top 25\% best fits). It can be seen from the table, the number of models within this range varies from 1 model for Ca\,{\sc ii} to 7 models for He\,{\sc i}.  Figure~\ref{fig:bd_FOMEllipse} illustrates where all these models fall in the explored parameter space of disk density $\rho_{0}$ and power law index $n$. If a single model is found, it is represented by a point. For two models, a line connecting the two models is shown on the figure. For three models, a triangle is used. For more than three models, an ellipse is shown that encloses most of the models. For He\,{\sc i}, a photospheric feature, the region that represents the models that reproduce no disk emission is shown with an arrow. It is important to keep in mind that this figure represents only the value of the power law index $n$ and disk density $\rho_{0}$; the rest of the parameters for the models ($R_{disk}$ and $i$) are not distinguished. As the figure illustrates, some, but not all, of the models overlap, again illustrating that no common region is found where all of the lines can be well-fit by a single power-law model. However, two general regions on the plot can be separated, one for the Balmer lines, which require relatively low densities and another region of higher densities, dominated by the metal lines Fe\,{\sc ii} and Ca\,{\sc ii}. This figure confirms the earlier observation that the metal lines require higher densities. 

\begin{table}
\caption {Number of models within 25\% of the least value of the figure of merit ${\cal F}_{min} \leq {\cal F} \leq 1.25\,{\cal F}_{min}$ using the global disk density model of Table~\ref{tab:bestfitmodels}.}
\label{tab:25FOM_models}
\smallskip
\begin{center}{\small}
   \begin{tabular}{ll}
   \hline
\hline
   \noalign{\smallskip}
  	Emission Line & Number of Models  \\ 
  	\hline
 	H$\alpha$ & 4\\ \noalign{\smallskip}
	H$\beta$ & 7 \\ \noalign{\smallskip}
	Ca\,{\sc ii} ($\lambda\,8542$) & 1\\ \noalign{\smallskip}
	Fe\,{\sc ii} ($\lambda\,5169$) &3 \\ \noalign{\smallskip}
	Fe\,{\sc ii} ($\lambda\,5317$) & 3\\ \noalign{\smallskip}
	%He\, {\sc i} ($\lambda\,6678$) & 38 \\ \noalign{\smallskip}
	\hline
	\noalign{\smallskip}
	\end{tabular}
	\end{center}
%\tablecomments{}
\end{table}

\begin{figure*}[h]
\centering

\begin{subfigure}[b]{0.99\textwidth}
\centering
\epsscale{1.0}
\plotone{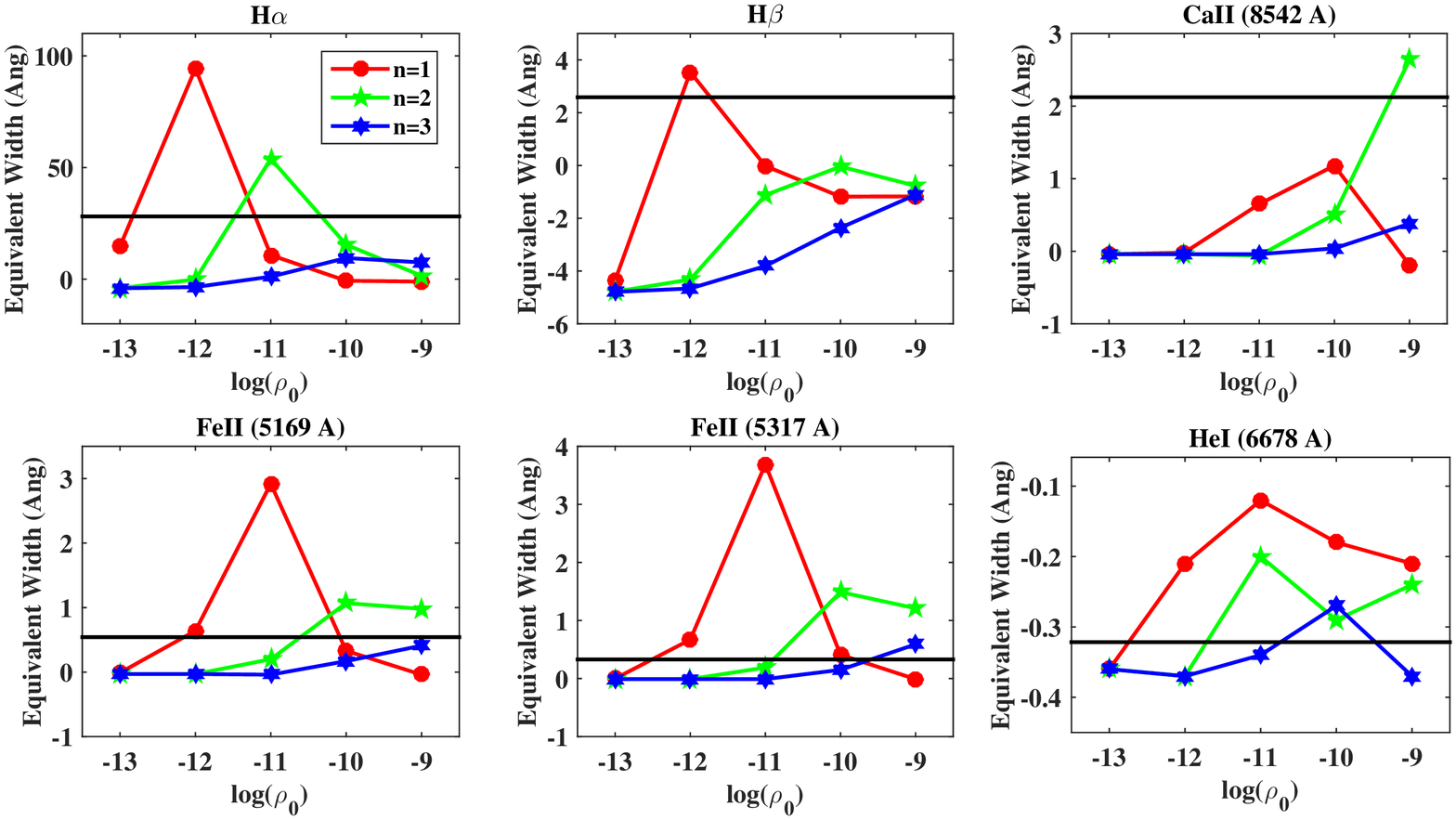}
\caption{\textit{Thin} disk model}
\label{fig:bd_EWvslogRho_thindisk}
\end{subfigure}

\begin{subfigure}[b]{0.99\textwidth}
\centering
\epsscale{1.0}
\plotone{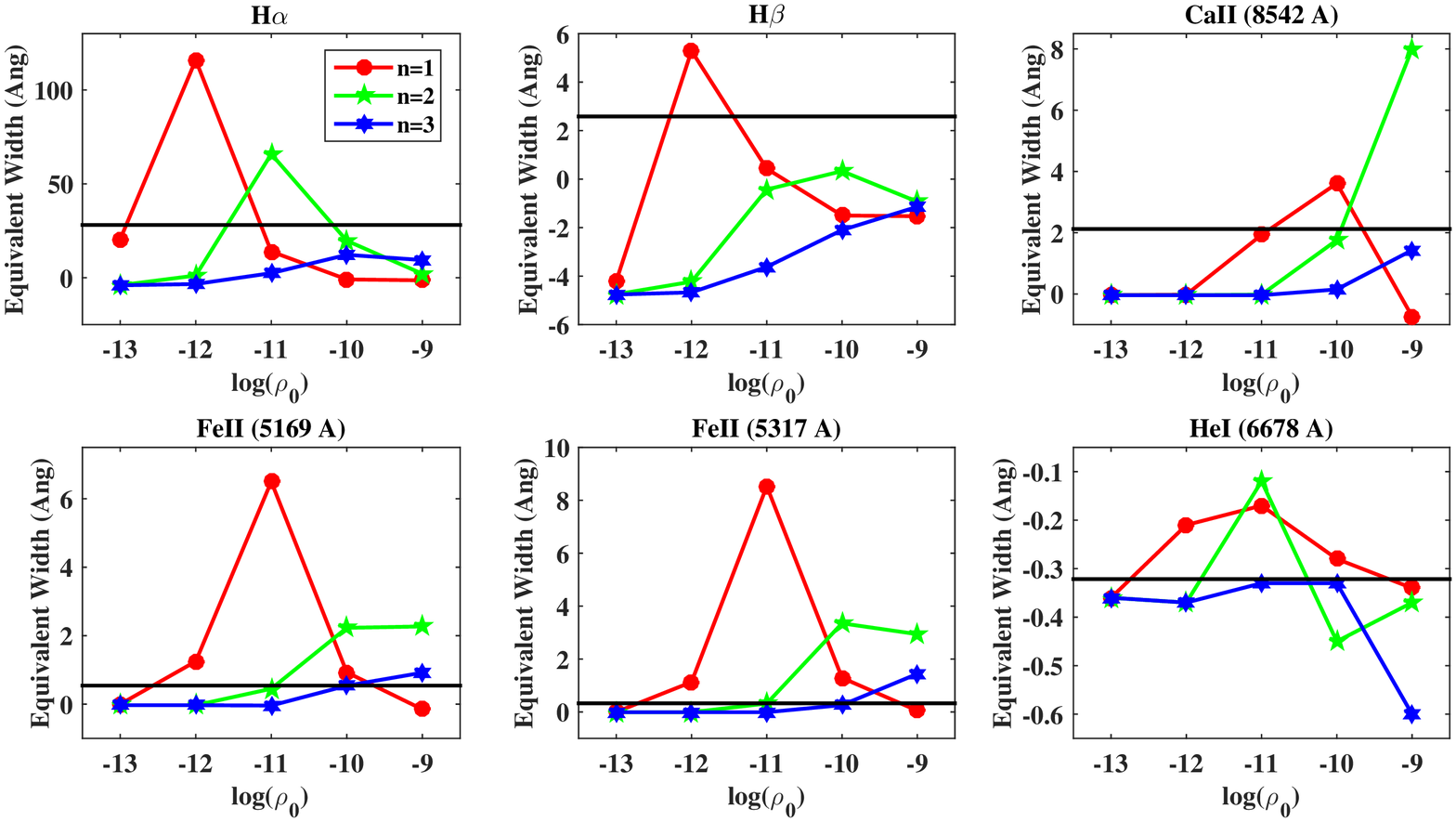}
\caption{\textit{Thin \& Turbulent} disk model}
\label{fig:bd_EWvslogRho_thinturbulentdisk}
\end{subfigure}

\caption{Equivalent width (\AA) of each individual line as a function of density (log($\rho_{0}$)) for three different power law indices $n$ for models of radius 50\,R$_{*}$ seen at 45$^{\circ}$ inclination angle. The solid black line in each panel indicates the respective observed equivalent width.}
\label{fig:bd_EWvslogRho_thin}
\end{figure*}

\begin{figure*}[h]
\centering

\begin{subfigure}[b]{0.99\textwidth}
\centering
\epsscale{1.0}
\plotone{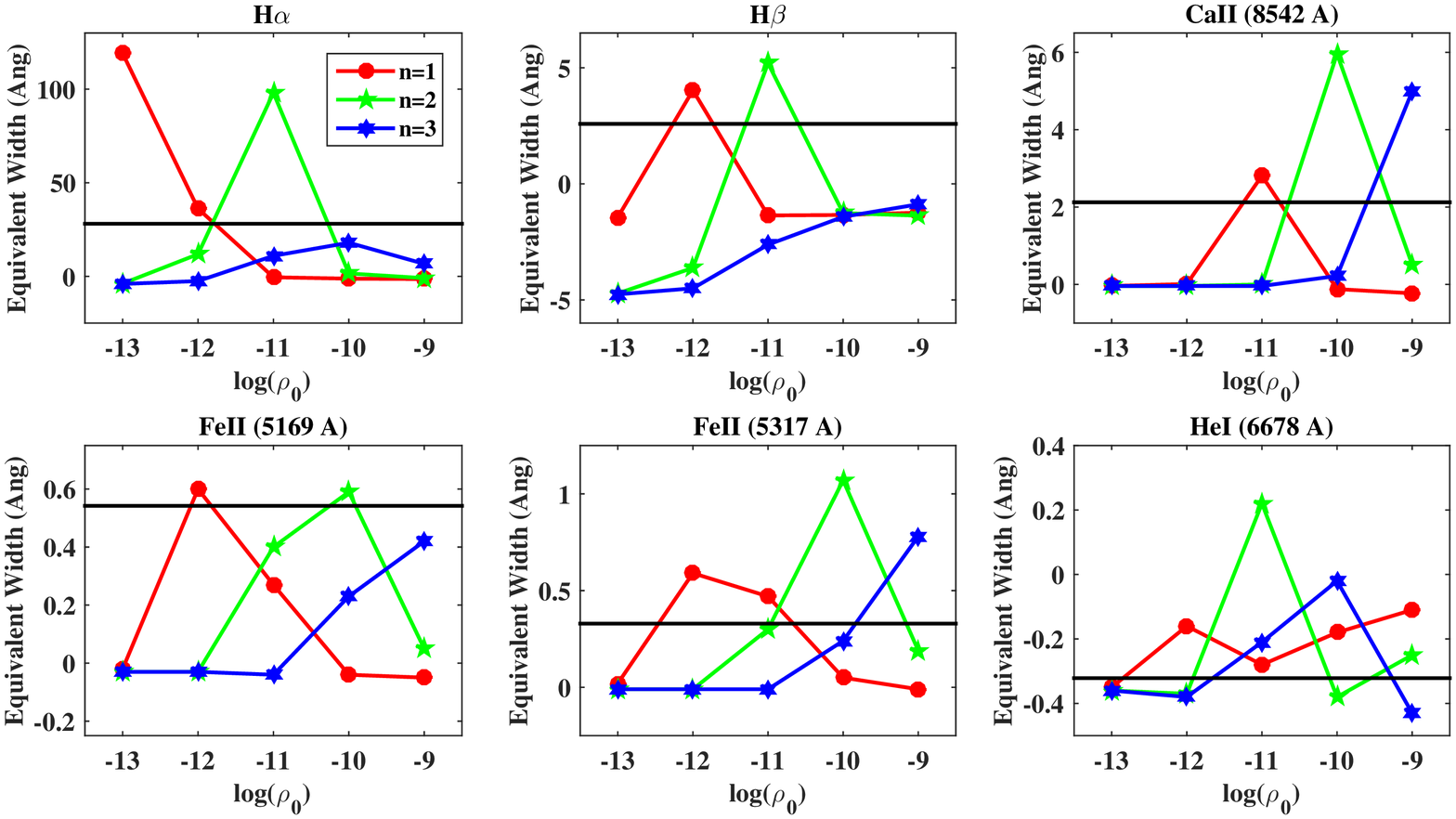}
\caption{\textit{Thick} disk model}
\label{fig:bd_EWvslogRho_thickdisk}
\end{subfigure}

\begin{subfigure}[b]{0.99\textwidth}
\centering
\epsscale{1.0}
\plotone{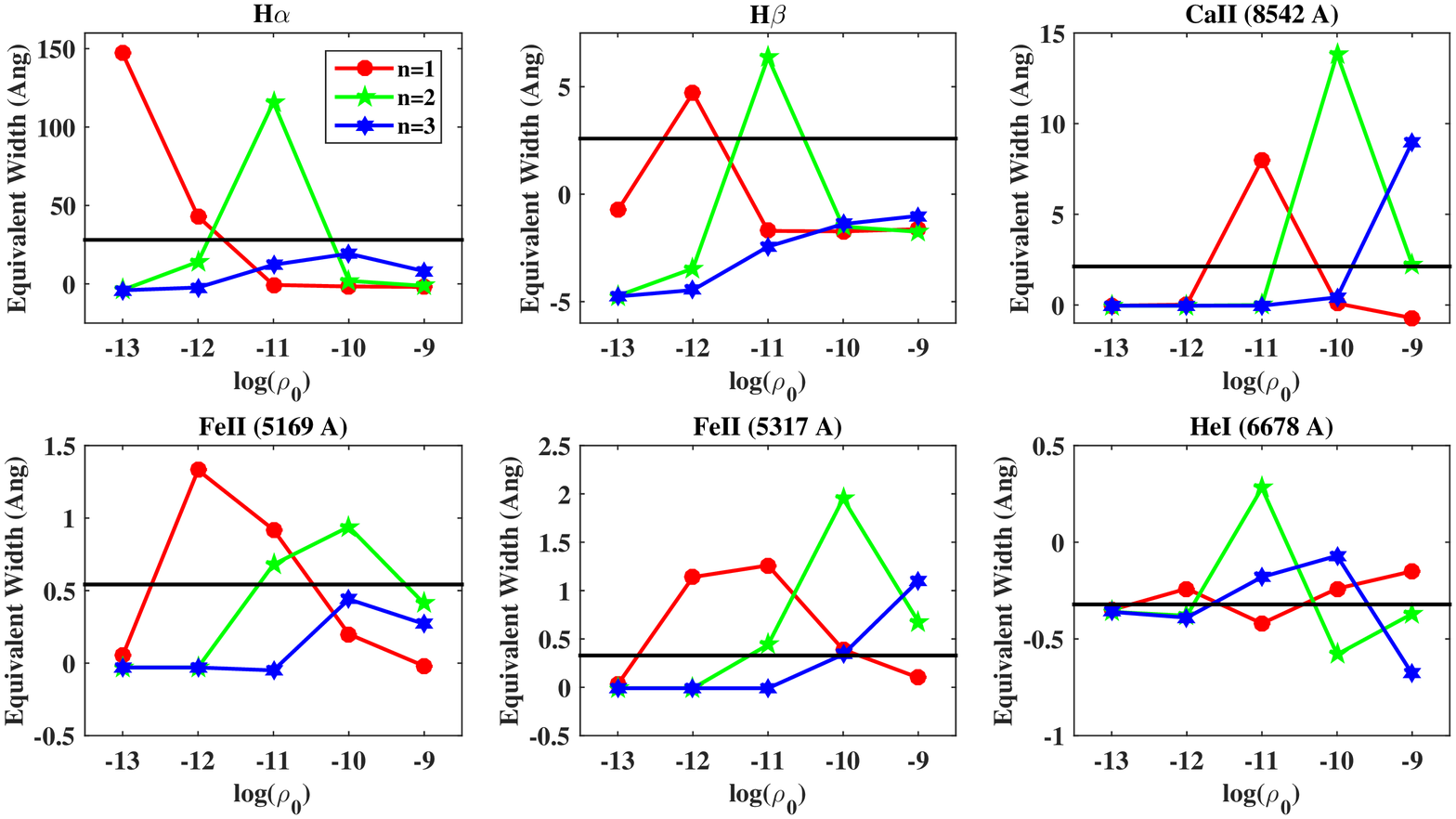}
\caption{\textit{Thick \& Turbulent} disk model}
\label{fig:bd_EWvslogRho_thickturbulentdisk}
\end{subfigure}

\caption{Equivalent width (\AA) of each individual line as a function of density (log($\rho_{0}$)) for three different power law indices $n$ for models of radius 50\,R$_{*}$ seen at 45$^{\circ}$ inclination angle. The solid black line in each panel indicates the respective observed equivalent width.}
\label{fig:bd_EWvslogRho_thick}

\end{figure*}

Given that a single model is not able to reproduce the observed line profiles, understanding how different power law indices and disk densities for the disk affect the overall line strengths is important. To this end, the EW for each line as a function of disk density parameter $\rho_{0}$ was plotted for models with a 50\,R$_{*}$ disk size seen at an inclination of 45$^{\circ}$. 
Figures~\ref{fig:bd_EWvslogRho_thin} and~\ref{fig:bd_EWvslogRho_thick} show the results for the four disk types considered here: \textit{thin, thick, thin and turbulent, and thick and turbulent}. Models with disk density parameter $\rho_{0}$ of $10^{-13}$, $10^{-12}$, $10^{-11}$, $10^{-10}$ and $10^{-9}~\rm g\,cm^{-3}$, and power law indices $n$ of 1, 2, and 3 are shown. In each figure, the black line indicates the observed EW for that particular emission line in the CFHT ESPaDOnS spectrum from 2006. 

Even as no single ($\rho_{0}$,n) combination is able to match the observed EW of all the lines, some models match the observed EW for more than one line. For example, in Figure~\ref{fig:bd_EWvslogRho_thinturbulentdisk} for a thin and turbulent disk, the EW of two Fe\,{\sc ii} lines and He\,{\sc i} line match the observed EW for the disk density $\rho_{0}$ of $10^{-10}~\rm g\,cm^{-3}$ and power law index $n$ of 3. The EW of H$\alpha$ for the same model is very close to the observed EW. However, the EW of H$\beta$ and Ca\,{\sc ii} ($\lambda\,8542$) are weaker for the same model when compared to the observations.

In general, the EWs increase with disk density $\rho_{0}$ to a maximum value, and then decline as the lines become saturated while the continuum continues to increase, weakening the EW. A good example of this can be seen for the power law index $n$ of 1, where a sharp increase to a peak and than decline can be clearly seen for each line. In general, the maximum EW moves to a higher disk density $\rho_{0}$ as the power law index $n$ increases. Also illustrated in the plots, the addition of turbulence increases the strength of the lines. The thicker disk models generally show a large number of models with EW equal or greater than the observed EW for all the lines. For example, for a power law index $n$ of 2, the models with thicker disks show stronger EWs and a stronger rise in the EW as the density increases. 

\section{Discussion}
\label{sec:discussion}

Good matches for all of the observed individual emission line profiles for BD+65\,1637 have been found in the large library of synthetic models. However, the diversity of the models in Table~\ref{tab:bestfitmodels}, and the failure to find one global model that fits all the observed line profiles well, seems to indicate that the density distribution within the inner gaseous disk of BD+65\,1637 cannot be of the simple form of a single power law (Equation~\ref{eq:rho}) with power law index $n>0$. The differences between the best-fits for individual lines and global fits suggest that the structure of the disk is more complex than a single power law. The idea that different density structure might be at play is supported by Figure~\ref{fig:bd_bestlinematches_cumfluxvsr} which illustrates how the variations in the structure of the disk can produce all the emission lines. In addition, the metal emission lines (Ca\,{\sc ii} and Fe\,{\sc ii}) seem to require a denser region for their formation as compared to the Balmer lines. 

The SED of the best-fit, global model overpredicts the near-IR excess compared to the available observations. However, it is important to note that SED is very sensitive to the underlying (assumed) stellar temperature in the optical/NIR, and hence should be viewed with caution. In general, the line modeling serves as a more powerful tool in inferring the structure of gaseous disk found close to the star. The comparison between the observed and computed SED was performed merely as a byproduct of the line modeling study performed here, and perhaps most importantly, the SED and line spectra observations are separated by 16 years.

A general trend was noticed while manually searching for the best-fit to the line profiles that the metal lines required higher densities compared to the Balmer lines in order to reproduce the observed line profiles. The addition of turbulence to these models made the lines stronger and broader. 

Finally, the analysis of observed and synthetic line profiles and their fits suggest that BD+65\,1637 is seen at an angle between 45$^{\circ}$ and 60$^{\circ}$.

Decretion disks around Classical Be stars are generally modeled with a single power law for the density structure, as mentioned in Section~\ref{sec:modeling} and~\cite{SJ2007}. When hydrodynamic models are used, a more complex density structure is predicted~\citep{Carciofi2011}. As shown by the current work, HAeBe stars do not seem to follow a single power law for their disk structure, perhaps as expected. Thus, a disk with density described by several different power laws in different radial zones might be able to provide a better global fit to all the lines considered. Finally, we assumed that the disk extends all the way to stellar photosphere, so another possible area of exploration would be to have the disk start further away from star. If the star isn't actively accreting, or has sporadic events of accretion, the disk may not extend all the way to the star. Many recent studies such as~\cite{Vink2015,Vink2005} have suggested that it may be possible to constrain the presence of such an inner hole radius using polarimetry. 

\section{Conclusions}
\label{sec:conclusions}
This study of inner gaseous disk of the Herbig B2e star, BD+65\,1637, by modeling the optical and near-infrared emission lines, has led to three key findings:
\begin{itemize}
\item All of the observed emission lines considered in this study can be reproduced with models that use \textit{photoionizing radiation of the central star} as the \textit{sole} energy source for the disk. 
\item Despite being able to reproduce the observed emission lines individually, no model based on a single power law for the equatorial density was able to reproduce all of the emission lines simultaneously. More complex density models are required to generate a consistent disk structure for this star.
\item The metal lines (Ca\,{\sc ii}, Fe\,{\sc ii}) require higher densities when compared to the Balmer lines. 
\end{itemize} 

In addition to testing more general disk density models, and investigating the effect of an inner hole in the disk, we have three more B2-type stars and four B0-type stars in the database~\citep{Alecian2013}. Applying the same modeling technique as in this work, we hope to further understand the overall geometry and structure of the disks around early-type HBe stars. For the next paper in this series, the B2-type stars, HD 76534, HD 216629 and HD 114981 will be analyzed.

\vspace{1.5\baselineskip}
Acknowledgments: This work is supported by the Canadian Natural Sciences and Engineering Research Council (NESRC) through Discovery Grants to T. A. A. Sigut and J. D. Landstreet.

\end{document}